\documentclass[aps,pra,amsmath,superscriptaddress,onecolumn,longbibliography,nofootinbib]{revtex4-2}

\usepackage{graphicx}
\usepackage{amsmath}
\usepackage{amsfonts}

\usepackage{amssymb}
\usepackage{mathrsfs}
\usepackage{braket}
\usepackage{bm}
\usepackage{multirow}
\usepackage{color}
\usepackage[normalem]{ulem}
\usepackage{mathrsfs}
\usepackage{mathtools}
\usepackage{float}
\usepackage{enumitem}
\usepackage{bbold}

\makeatletter

\newcommand{\Rmnum}[1]{\expandafter\@slowromancap\romannumeral #1@}
\makeatother

\def\ket#1{|\, #1\,\rangle}
\def\bra#1{\langle\,#1\,|}

\def\tr{{\rm tr}\,}

\def \sx{\hat\sigma^{\rm x}}
\def \sz{\hat\sigma^{\rm z}}
\def \sy{\hat\sigma^{\rm y}}

\def\S{\hat {\cal S}}
\def\openone{{1\!\!1}}

\begin{document}

\title{Fast pseudorandom quantum state generators via inflationary quantum gates}

\author{Claudio Chamon}
\email[Corresponding author: ]{chamon@bu.edu}
\affiliation{Physics Department, Boston University, Boston, Massachusetts 02215, USA}

\author{Eduardo R. Mucciolo}
\affiliation{Department of Physics, University of Central Florida, Orlando, Florida 32816, USA}

\author{Andrei E. Ruckenstein}
\email[Corresponding author: ]{andreir@bu.edu}
\affiliation{Physics Department, Boston University, Boston, Massachusetts 02215, USA}

\author{Zhi-Cheng Yang}
\affiliation{School of Physics, Peking University, Beijing 100871, China}
\affiliation{Center for High Energy Physics, Peking University, Beijing 100871, China}

\date{\today}

\begin{abstract}

We propose a mechanism for reaching pseudorandom quantum states, computationally indistinguishable from Haar random, with shallow log-n depth quantum circuits, where n is the number of qudits. We argue that $\log n$ depth 2-qubit-gate-based generic random quantum circuits that are claimed to provide a lower bound on the speed of information scrambling, cannot produce computationally pseudorandom quantum states. This conclusion is connected with the presence of polynomial (in n) tails in the stay probability of short Pauli strings that survive evolution through such shallow circuits. We show, however, that stay-probability-tails can be eliminated and pseudorandom quantum states can be accomplished with shallow $\log n$ depth circuits built from a special universal family of `inflationary' quantum (IQ) gates. We prove that IQ-gates cannot be implemented with 2-qubit gates, but can be realized either as a subset of
 2-qudit-gates in $U(d^2)$ with $d\ge 3$ and $d$ prime, or as special 3-qubit gates. 

\end{abstract}

\maketitle

\section*{Introduction}
\label{sec:intro}

The focus of this paper is on addressing the following question: what
is the lowest-depth quantum circuit that can generate computationally
pseudorandom quantum states, i.e., quantum states that cannot be
distinguished from Haar random by an adversary limited to polynomial
resources? For us this question was motivated by two conjectures
concerning scrambling by quantum circuit models of black hole dynamics
that have emerged in the context of decades-old efforts of reconciling
general relativity with quantum mechanics. The first, due to Susskind
and collaborators, is that black holes are the fastest scramblers in
nature with a scrambling time $\tau _{sc} \sim \log n$, where $n$ is
the number of degrees of freedom of the
system~\cite{Sekino-Susskind2008}, a conjecture supported by
holography-based
calculations~\cite{Shenker2014,Shenker2014b,Maldacena2016}.  The
second conjecture is that black holes must be also thorough scramblers
of
information~\cite{Hayden-Harlow2013,Kim-Tang-Preskill2020,Bouland-Fefferman-Vazirani2019}. In
a nutshell, the idea is that black holes are also efficient generators
of (computational) pseudorandomness: that they scramble information
efficiently (i.e., in polynomial time) but that unscrambling
(decoding) that information requires superpolynomial (in $n$)
effort. While convincing arguments have been advanced for each of
these conjectures, the question of whether one can achieve both
`speed' and `thoroughness' at the same time - namely whether a
quantum circuit of $\log n$ depth (corresponding to a `computational
time' scaling as $\log n$) can create pseudorandom states
indistinguishable from Haar random for an adversary with polynomial
resources - has, to our knowledge, not been discussed explicitly in
the black hole literature.

Irrespective of whether or not satisfying both `speed' and
`thoroughness' conditions is critical to understanding the quantum
mechanics of black holes, the question of the level of scrambling by
$\log n$-depth circuits is conceptually and practically important to a
number of areas of quantum information. In particular, the issue has
been discussed in the context of $t$-designs. In their comprehensive
studies, Harrow and Mehraban~\cite{Harrow2018approximate} conjectured
that complete-graph-structured $\log n$-depth random circuits display
the property of anti-concentration, namely that the probability that
two different realizations of the circuit lead to identical outcomes
is exponentially small in $n$, and at most a constant multiple of the
value obtained by averaging over the Haar measure. They also
conjectured that $\log n$-depth long-range circuits are sufficient for
reaching 2-designs. While the anti-concentration conjecture was
recently proved for a few circuit
connectivities, anti-concentration is not
sufficient for proving 2-design~\cite{Barak2020,Brandao2022}. In particular, we show below that
$\log n$-depth long-range 2-qubit circuits lead to polynomial (rather
than superpolynomial) decay of a 4-point out-of-time-order correlator,
and therefore these shallow circuits cannot produce 2-designs. We note
that being a 2-design is a stronger (statistical) condition than
computational indistinguishability based on measurements of
2-time/4-operator correlations. The notion of $t$-design
  refers to `statistical indistinguishability' from Haar-random,
  which is determined using the distance between probability
  distributions or the difference between the correlations they
  produce. By contrast, the discussions of pseudorandomness and all
  arguments of this paper are limited to `computational
    indistinguishability', a more physical notion referring to
  adversaries who are limited to a polynomial number of measurements.
Building computational pseudoramdom Boolean functions with
$\log n$-depth (NC$^1$) circuits, a closely related classical version
of the question we ask of quantum circuits, has been addressed by the
cryptography
community~\cite{Naor1997NumbertheoreticCO,Naor2002,Applebaum2016}. These
classical constructions, however, involve pre-processing and
non-trivial storage considerations that are not obviously amenable to
low-depth quantum implementations.

Our own interest in information scrambling and the issues raised in
this paper stem from our work on $n$-input/$n$-output
reversible-circuit-based classical block ciphers and, in particular,
on the question of what is the fastest, lowest-depth block
cipher that is secure to attacks by polynomially-limited
adversaries. In Ref.~\cite{cipher-paper} we proposed a cipher design
that is capable of scrambling information with only ${\cal O}(\log n)$
layers of gates, on a par with the conjectured fastest scrambling time
by black holes, but, we argued, to cryptographic level: the special
$\log n$-depth cipher produces a permutation which is computationally
indistinguishable from pseudorandom to an adversary with polynomial
resources. [We stress that the ${\cal O}(\log n)$-depth
  cipher design in Ref.~\cite{cipher-paper} meets necessary
  conditions for indistinguishability from a pseudorandom
  permutation. These conditions are based on quantitative measures of
  chaos and irreversibility in quantum systems (such as
  out-of-time-order correlators and string entropies). We do not
  establish sufficiency of these measures as this would be equivalent
  to proving that P$\ne$ NP.] It is also worth noting that these ciphers are NC$^1$
reversible circuits, implemented without the need for
preprocessing or additional storage~\cite{Naor1997NumbertheoreticCO,Naor2002,Applebaum2016}.

At first sight, classical ciphers seem only distantly related to the
problem of information scrambling by quantum circuits.  However, our
progress in designing fast classical ciphers was based on a mapping of
reversible classical computations into the space of Pauli
strings. Within the framework of strings, the notions of
irreversibility and chaos and their quantitative measure in terms of
string entropies and out-of-time-order correlators (OTOCs) used in
studies of quantum scrambling translate naturally to the problem of
scrambling by reversible classical circuits. It is the string space
picture that allows us to use the intuition gained from the study of
one problem to the study of the other.


\begin{figure*}
\centering
\includegraphics[width=0.8\textwidth]{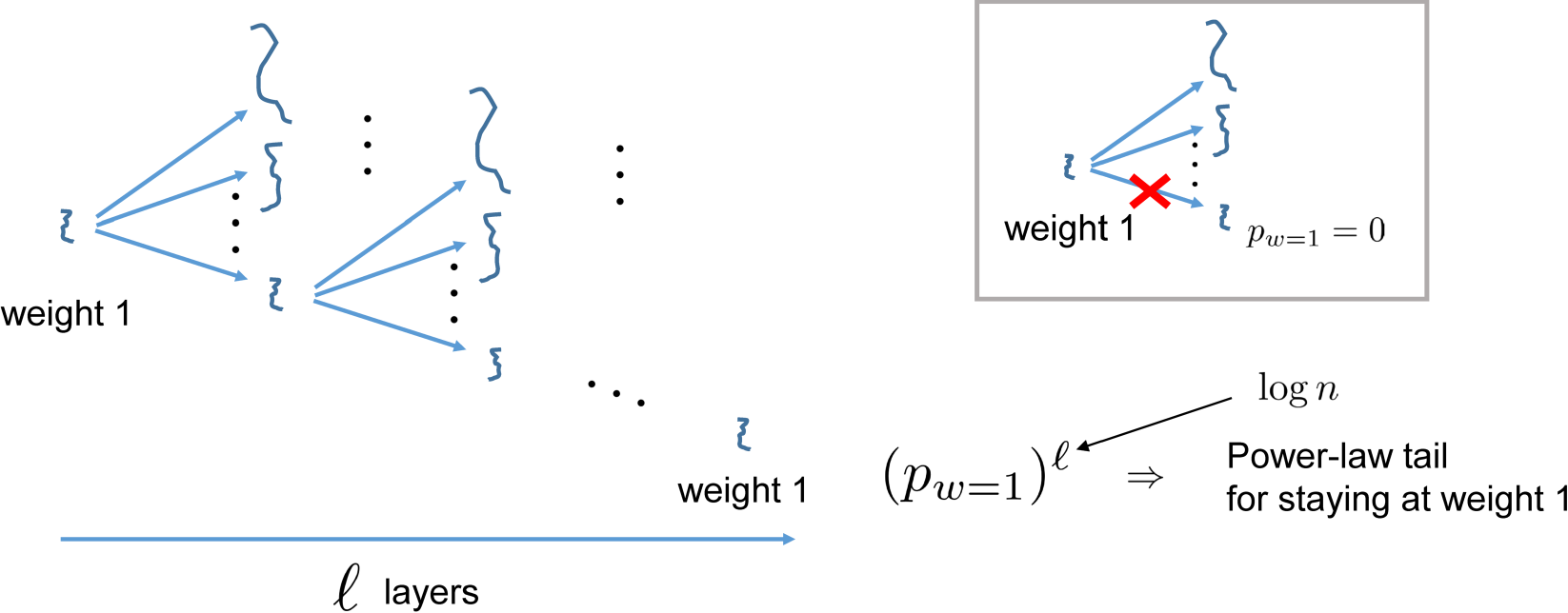}
\caption{{\bf Evolution of weight-1 Pauli strings:} an effective 2-qubit
  reversible gate obtained by averaging uniformly over 2-qubit gates
  in $U(4)$ leads to equal transition amplitudes among the 15
  (=$4^2-1$) non-trivial weight-1 and weight-2 string states,
  including a finite stay probability, $p_{w=1} < 1$, for weight-1
  strings (i.e., a finite amplitude for the transition from a weight-1
  to another weight-1 string state). Applying $\log n$ layers of
  2-qubit gates leads to a polynomial tail
  $(p_{w=1})^{\log n}=n^{-{\log (1/p_{w=1})}}$ in the stay probability
  of weight-1 strings that, in turn, translates into polynomial tails
  in OTOCs. Exponential decay of OTOCs in $\log n$ depth quantum
  circuits requires the use of special `inflationary' gates that map
  all weight-$1$ strings into weight-$2$ strings, thus eliminating the
  stay probability for weight-1 strings, as depicted schematically in
  the inset.}
\label{fig:stay_prob}
\end{figure*}


In particular, in the context of random reversible classical circuits,
the repeated forward and backward propagation that defines OTOCs
involving a polynomial number of string operators naturally describes
arbitrary polynomial measurements carried out by an adversary on
inputs and/or outputs of the circuits. Security to such attacks -
referred to as differential attacks in cryptoanalysis - requires that
all OTOCs describing correlations between results of such alternating
measurements vanish faster than any polynomial (and ideally
exponentially) in the number of bits acted on by the circuit.
However, as discussed in Ref.~\cite{cipher-paper} for unstructured
classical random circuits of universal reversible gates and as
explained in the body of the paper for $2$-qubit-gate-based random
quantum circuits, the typical OTOC vanishes exponentially with
`computational time' - the number of layers of gates applied. (The
same exponential decay of the OTOC with time is the generic behavior
expected for scrambling of information and the approach to chaos in
quantum systems described by unitary Hamiltonian
evolution~\cite{Maldacena2016}.) As a result, generic circuits of
$\log n$ depth lead to polynomial decays of the OTOCs, and thus
are not secure to polynomial attacks (in the classical case) and do
not generate computationally pseudorandom quantum states (in the quantum case).

The first message of this paper is that generic random quantum
circuits of 2-qubit gates like those used as simple models of black
hole dynamics cannot produce computationally pseudorandom quantum
states while at the same time saturating the $\log n$ lower bound on
the scrambling time purported to qualify black holes as the fastest
scramblers in nature. The root of the problem is the presence of
polynomial tails of the stay-probabilities for low-weight Pauli
strings, illustrated schematically in Fig.~\ref{fig:stay_prob}. In
turn, for generic $\log n$-depth random circuits, these tails
translate into a polynomial decay of OTOCs with $n$. Using the
intuition gained from the study of shallow classical ciphers in
Ref.~\cite{cipher-paper}, we argue that ensuring a superpolynomial
decay of OTOCs in $\log n$-depth quantum circuits requires employing
special `inflationary gates' that eliminate the stay-probability of
weight-1 strings and accelerate the spreading of string operators (see
the inset of Fig.~\ref{fig:stay_prob}).

Our second message is that, while inflationary gates do not exist as
2-qubit gates in $U(4)$, they can be realized as 2-qudit gates
with local Hilbert space dimension $d\ge 3$ and $d$ prime, or as
$3$-qubit gates. Circuits built from these gate sets would implement
cryptographic level scrambling at the $\log n$ `speed limit,'
performance one might like to ascribe to a supreme `superscrambling'
black hole.

Finally, we note that, while saturating the $\log n$ scrambling time
lower bound may not be critical for resolving black hole paradoxes,
reaching computationally pseudorandom permutations at the $\log n$ `speed limit' for scrambling was
crucial in our own work on classical ciphers. As discussed in a recent
paper~\cite{EOC}, ciphers of $\log n$-depth enable Encrypted Operator
Computing (EOC), a gate-based polynomial complexity approach to
secure computation on encrypted data that offers an alternative to
Fully Homomorphic Encryption. For larger-depth circuits (and even for circuits of $\log ^2 n$ depth), the
implementation of the EOC scheme would become superpolynomial in $n$
and thus computationally intractable.

\section*{Results}

\subsection*{Our contributions}
\label{sec:contributions}

The principal conclusions of this paper are that:
\begin{itemize}
\item Due to polynomial tails in the stay-probability for weight-$1$
  Pauli strings, 2-qubit-gate-based random quantum circuits of depth
  ${\cal O}(\log n)$ cannot produce pseudorandom states.
\item Reaching pseudorandom quantum states with
  ${\cal O}(\log n)$-depth circuits becomes possible if one employs
  circuits comprised of special universal $3$-qubit gates or 2-qudit gates with
  local Hilbert space dimension $d\ge 3$ and $d$ prime. These gates,
  which we refer to as `inflationary quantum gates', or IQ gates, both
  expand and proliferate Pauli strings.
\end{itemize}
These conclusions are built on the intuition gained from our work in
Ref.~\cite{cipher-paper} on classical ciphers based on reversible
circuits of $\log n$ depth, which we translate to the problem of
information scrambling by quantum circuits. The polynomial tails in
the probability distribution of weight-$1$ strings also occur in
random reversible classical circuits and it is the elimination of
these tails that required the structured design of our $\log n$-depth
classical cipher. This design involves a
permutation $\hat P$ expressed as a $3$-stage circuit
$\hat P=\hat L_r\;\hat N\;\hat L_l$, with each stage represented by
tree-structured reversible classical circuits built out of 3-bit
permutations (gates), which enable universal classical computing. (The
wiring of tree-structured circuits is described in the `Tree-structured circuits' discussion in the Methods section.)
The bookends $\hat L_{l,r}$
are comprised of $\log_2 n$ layers of special (classical) linear
inflationary gates, that flip at least two output bits upon flipping a
single input. The implementation of inflationary gates in string space
eliminates the stay probability of weight-1 strings and accelerates
the spreading of their effect across the $n$ bitlines of the
circuit. These inflationary
stages flank a reversible circuit $\hat N$, comprised of $\log_3 n$
layers of (classical) nonlinear gates that maximize production of
(Pauli)-string entropy.  As argued in Ref.~\cite{cipher-paper}, the
3-stage circuit realizes a $\log n$-depth cipher that satisfies the
necessary (and we conjecture sufficient) conditions for
pseudorandomness. [We note that the tree structure mimics a system of infinite
  dimension and, when combined with the inflationary property of the
  gates, ensures that the weight of the strings grows exponentially
  with the depth or number of layers of gates.] 

The interplay between
inflation and proliferation of strings leads to a double exponential
decay of OTOCs as a function of the computational time, i.e., the
number of layers of gates. For our shallow $\log n$-depth cipher 
this double exponential behavior, which
implies an infinite Lyapunov exponent, translates into an
exponential decay of OTOCs with $n$. We note that, in classical
circuits, inflation and proliferation of strings are implemented by
different families of gates and thus, as described above, fast and
thorough scrambling requires a structured 3-stage cipher. In this
paper we exploit the interplay of inflation and proliferation of
strings in the context of quantum circuits. Unlike the case of
classical circuits, in the quantum case one can build IQ gates that
incorporate both string inflation and string proliferation. As a
result, fast and thorough quantum scrambling can be realized with
unstructured single-stage random quantum circuits comprised of IQ
gates.


\subsection*{Generating pseudorandom quantum states with $\log n$-depth circuits}
\label{sec:pseudorandom_states}

\begin{figure*}
\centering
\includegraphics[width=0.9\textwidth]{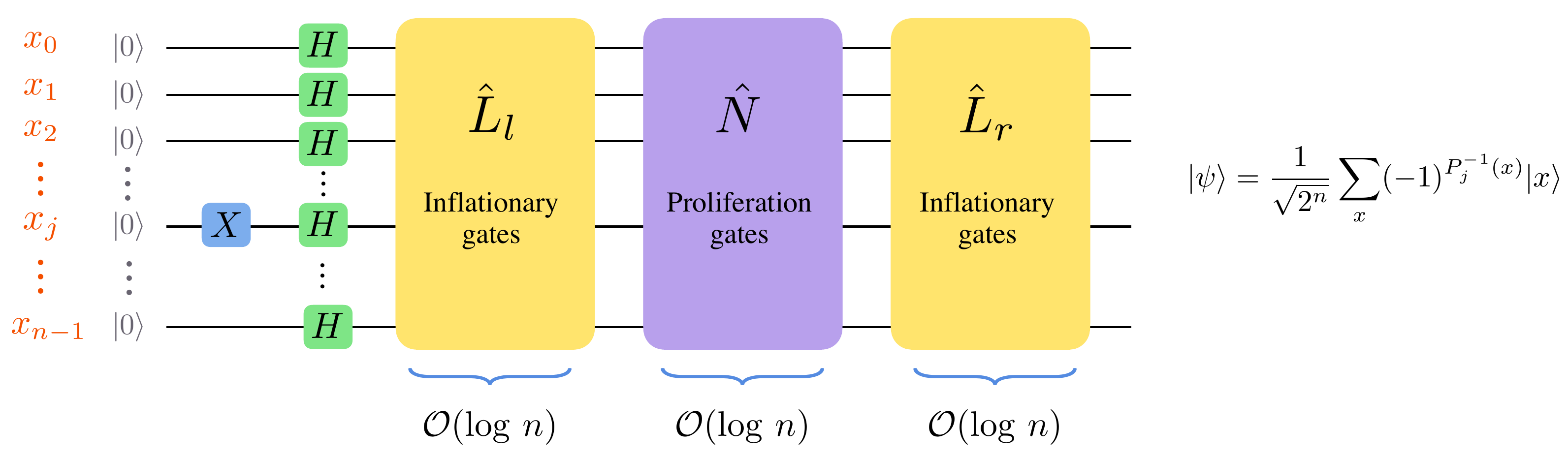}
\caption{{\bf The circuit architecture that generates the pseudorandom quantum state with depth $\mathcal{O}({\rm log}\ n)$.} This construction uses the three-stage (computationally) pseudorandom permutation of Ref.~\cite{cipher-paper}, which is built out of  three circuits, $\hat{L}_{l/r}$ and $\hat{N}$, comprised of linear inflationary gates and nonlinear proliferation gates, respectively.  Each block contains $\mathcal{O}(n \ {\rm log} n)$ three-qubit gates organized into a tree structure of $\mathcal{O}({\rm log} \ n)$ layers (see the `Tree-structured circuits' discussion in the Methods section.)}
\label{fig:pseudorandom}
\end{figure*}

In this section we first argue that $\log n$-depth random quantum
circuits built by sampling uniformly over 2-qubit gates in $U(4)$
cannot produce pseudorandomness. We
then give an example, schematically depicted in
Fig.~\ref{fig:pseudorandom}, of how to construct a pseudorandom state
by employing a quantum circuit comprising a layer of Hadamard gates
followed by the $\log n$-depth 3-stage classical cipher of
Ref.~\cite{cipher-paper} and described in the previous section.
Sec.~\ref{sec:contributions}. Given that the classical cipher produces
a pseudorandom permutation, an assumption tested via the Strict
Avalanche Criterion (SAC) for pseudorandomness of classical
ciphers~\cite{Feistel,Llyod,Hirose1995}, we show that the resulting
quantum state satisfies the pseudorandomness condition expressed in Eq.~\eqref{eq:pseudorandom-condition} below. 

We proceed by relating quantum expectation values of string operators
to OTOCs, an identity which turns out to be useful in establishing the
results of this section. We consider a quantum state $\ket{\psi}$ on
the Hilbert space of $n$ qubits that is obtained through the evolution
via a unitary transformation $\hat U$ applied to an initial product
state $\ket{\psi_0}$: $\ket{\psi}= \hat U\,\ket{\psi_0}$.  If the
state $\ket{\psi}$ is pseudorandom, then the expectation values of
(non-trivial) Pauli string operators must vanish faster than any
polynomially bounded function $\eta(n)$ of the number of qubits, $n$:
\begin{align}
  \label{eq:pseudorandom-condition}
  |\bra{\psi}\;\S_\alpha\;\ket{\psi}|^2 < \eta(n)
\;,
\end{align}
where a Pauli string,
\begin{align}
  \label{eq:string-alpha}
  \S_\alpha
  =
  \prod_{j\in\alpha^{\rm x}}\;\sx_j
  \;
  \prod_{k\in\alpha^{\rm z}}\;\sz_k
  \;,
\end{align}
is labeled by the set $\alpha=(\alpha^{\rm x},\alpha^{\rm z})$ of
qubit indices present in the string. By adding a phase
$i^{\alpha^{\rm x}\cdot\alpha^{\rm z}}$ to $\S_\alpha$ -- picking up
an $i$ each time both a $\sx_j$ and $\sz_j$ appear at the same $j$, or
basically deploying the $\sy$s as well -- would make the string
operator Hermitian. Here we prefer the definition
Eq.~(\ref{eq:string-alpha}) for the applications we consider, and work
explicitly with both $\S^{\; }_\alpha$ and $\S^{\dagger}_\alpha$ when
needed. (When convenient, we also use the equivalent notation
$\alpha^{\rm x,z}_i=1 \leftrightarrow i\in \alpha^{\rm x,z}$ and
$\alpha^{\rm x,z}_i=0 \leftrightarrow i\notin \alpha^{\rm x,z}$, as
for example in the definition of the dot product $a\cdot b \equiv
\sum_i a_i b_i$.)

We next move the unitary transformation onto the string operators to
rewrite the left hand side of Eq.~\eqref{eq:pseudorandom-condition} in
terms of `time'-evolved Pauli strings, $\S_\alpha(\tau)\equiv\hat
U^\dagger\;\S_\alpha\;\hat U$, and $\S_\alpha(-\tau)\equiv\hat
U\;\S_\alpha\;\hat U^\dagger$ ($\S_\alpha(0)\equiv\S_\alpha$).
Without loss of generality, we consider an initial product state in
the computational basis, $\ket{\psi_0}=\ket{x}$, where $x$ is an
$n$-bit binary vector. We can then write:
\begin{align}
  \label{eq:expectation-squared-one-x}
  &\left|\bra{x}\;\S_\alpha(\tau)\;\ket{x}\right|^2
  =
    \tr\left[\hat{\cal P}_x\;\S^\dagger_\alpha(\tau)\;\hat{\cal P}_x\;\S_\alpha(\tau)\right]
    \nonumber\\
  &=
    \frac{1}{4^n}\sum_{\beta^{\rm z},\beta'^{\rm z}}\;(-1)^{(\beta^{\rm z}\oplus\beta'^{\rm z})\cdot x}\;
    \tr\left[\S_{\beta^{\rm z}}\;\S^\dagger_\alpha(\tau)\;\S_{\beta'^{\rm z}}\;\S_\alpha(\tau)\right]
  \;,
\end{align}
where $\hat{\cal P}_x$ is the projector onto the state $\ket{x}$,
expressed in terms of Pauli strings as:
\begin{align}
  \hat{\cal P}_x = \prod_i \left[\frac{1+(-1)^{x_i}\sigma^z_i}{2}\right]
  =
  \frac{1}{2^n}\sum_{\beta^{\rm z}} (-1)^{\beta^{\rm z}\cdot x}\;\S_{\beta^{\rm z}}
  \;.
\end{align}
\begin{widetext}
If the correlator in Eq.~\eqref{eq:expectation-squared-one-x} decays
superpolynomially in $n$ for all initial states $\ket{x}$, then the
superpolynomial decay carries over to the average over all initial
states. Thus, averaging Eq.~\eqref{eq:expectation-squared-one-x} over
$x$, and using $\sum_x (-1)^{(\beta^{\rm z}\oplus\beta'^{\rm z})\cdot
  x}=2^n\,\delta_{\beta^{\rm z},\beta'^{\rm z}}$, yields
\begin{align}
\label{eq:Q}
  Q_\alpha(\tau)\equiv
  &\,\frac{1}{2^n}\,\sum_x\;\left|\bra{x}\;\S_\alpha(\tau)\;\ket{x}\right|^2
    \nonumber\\
  =&\,
     \frac{1}{2^n}\sum_{\beta^{\rm z}}\;
     \left\{\frac{1}{2^n}\;\tr\left[\S_{\beta^{\rm z}}\;\S^\dagger_\alpha(\tau)\;\S_{\beta^{\rm z}}\;\S_\alpha(\tau)\right]\right\}
     \nonumber\\
  =&\,\frac{1}{2^n}\sum_{\beta^{\rm z}}\;
    \left\{\frac{1}{2^n}\;\tr\left[\S^\dagger_{\beta^{\rm z}}(-\tau)\;\S^\dagger_\alpha\;\S_{\beta^{\rm z}}(-\tau)\;\S_\alpha\right]\right\}
                   \;,
\end{align}
where we shifted the time-dependence from $\tau$ to $-\tau$ by using
the cyclic property of the trace and the fact that the $z$-string
operator $\S_{\beta^{\rm z}}$ is Hermitian. Notice that the expression
within parentheses in Eq.~\eqref{eq:Q} represents an OTOC of Pauli
string operators.

Eq.~\eqref{eq:Q} can be translated into a more intuitive form by
writing the $\tau$-dependent string, $U\, \S_\alpha \,U^\dagger =
\sum_\beta\;A_{\alpha\beta}(-\tau)\;\S_\beta$ in terms of string
amplitudes, $A_{\alpha\beta}(-\tau)$, and then expressing
$Q_\alpha(\tau)$ as
\begin{align}
  Q_\alpha(\tau)=
  &\;\frac{1}{2^n}\sum_{\beta^{\rm z}}\;\sum_{\gamma,\gamma'}\;
    A^*_{\gamma\beta^{\rm z}}(-\tau)\;A_{\gamma'\beta^{\rm z}}(-\tau)\;\left\{\frac{1}{2^n}\;\tr\left[\S^\dagger_{\gamma}\;\S^\dagger_\alpha\;\S_{\gamma'}\;\S_\alpha\right]\right\}
    \nonumber\\
  =&\;\frac{1}{2^n}\sum_{\beta^{\rm z}}\;\sum_{\gamma,\gamma'}
    A^*_{\gamma\beta^{\rm z}}(-\tau)\;A_{\gamma'\beta^{\rm z}}(-\tau)\;\delta_{\gamma,\gamma'}\;(-1)^{\alpha^{\rm x}\cdot\gamma^{\rm z}}\;(-1)^{\alpha^{\rm z}\cdot\gamma^{\rm x}}
    \nonumber\\
  =&\;\frac{1}{2^n}\sum_{\beta^{\rm z}}\;\sum_{\gamma}
    \;|A_{\gamma\beta^{\rm z}}(-\tau)|^2\;\;(-1)^{\alpha^{\rm x}\cdot\gamma^{\rm z}}\;(-1)^{\alpha^{\rm z}\cdot\gamma^{\rm x}}
    \;.
\end{align}
For simplicity we consider a local $z$-string, $\S_\alpha=\sz_i$ (i.e.,
$\alpha^{\rm x}=0,\alpha^{\rm z}_j=\delta_{ij}$), in which case,
$Q_{\sz_i} (\tau)\equiv \frac{1}{2^n}\,\sum_x
|\bra{x}\;\sz_i(\tau)\;\ket{x}|^2$ is given by
\begin{align}
\label{eq:Qzi}
  Q_{\sz_i} (\tau)
  &=\frac{1}{2^n}\sum_{\beta^{\rm z}}\;\left[
    \sum_{\gamma|\sx_i\notin\S_\gamma}
    \;|A_{\gamma\beta^{\rm z}}(-\tau)|^2 -
    \sum_{\gamma|\sx_i\in\S_\gamma}
    \;|A_{\gamma\beta^{\rm z}}(-\tau)|^2
    \right]
    \nonumber\\
  &=\frac{1}{2^n}\sum_{\beta^{\rm z}}\;\left[
    (p_{i;\openone} (-\tau;\beta^{\rm z})+p_{i;z}(-\tau;\beta^{\rm z}))-(p_{i;x}(-\tau;\beta^{\rm z})+p_{i;y}(-\tau;\beta^{\rm z}))
    \right]    \;,
\end{align}
where
$p_{i;\openone}(-\tau;\beta^{\rm z}),p_{i;z} (-\tau;\beta^{\rm z}),p_{i;x}(-\tau;\beta^{\rm
    z}$ and $p_{i;y} (-\tau;\beta^{\rm z})$ are the probabilities that, at
position $i$ (and computational time $-\tau$), the Pauli string contains, respectively, an identity, a
$\sz$, a $\sx$, or the product $\sx\sz$. [Throughout we keep track of  the initial non-trivial string state
$\beta ^{\rm z}$ defining the transition amplitudes $A_{\gamma\beta^{\rm z}}(-\tau)$ in
Eq.~\eqref{eq:Qzi}.]

\end{widetext}

\noindent{\bf {Unstructured random quantum circuits:}}
\label{sec:unstructured}
We will now make use of Eq.~\eqref{eq:Qzi} to address the following
question: can a $\log n$-depth random quantum circuit built from
2-qubit gates (2-local) in $U(4)$ representing the unitary operator
$\hat U$ in $\ket{\psi}= \hat U\,\ket{\psi_0}$ lead to pseudorandom
states for which $Q_{\sz_i}$ satisfies the pseudorandomness condition
in Eq.~\eqref{eq:pseudorandom-condition}? We answer this question by
considering the average string weight, which is obtained by averaging
Eq.~\eqref{eq:Qzi} uniformly over circuits and over the axis of
quantization, whereby we can write
$\overline{p_{i;x}(-\tau;\beta^{\rm z})}
=\overline{p_{i;y}(-\tau;\beta^{\rm
    z})}=\overline{p_{i;z}(-\tau;\beta^{\rm z})}=\frac{1}{3} \rho (-\tau;\beta^{\rm z})$, where $\rho(-\tau;\beta^{\rm z})$ is the string density. As a result,
\begin{align}
\label{eq:Qaverage}
  \overline{Q_{\sz}(\tau)}
  &=\frac{1}{2^n}\sum_{\beta^{\rm z}}\;
    \left[
    1-\frac{4}{3}\; \rho(-\tau;\beta^{\rm z})
    \right]
    \;.
\end{align}
The averaging in Eq.~\eqref{eq:Qaverage} is carried out over random
$2$-qubit-gate-based universal circuits in which case the
superpolynomial bound on $Q_{\sz_i}(\tau)$ for a given random circuit
remains valid for the average $\overline{Q_{\sz}(\tau)}$. Here we
assume that ${Q_{\sz _i}(\tau)}$ obtained for a typical random circuit
coincides with the average over circuits,
$\overline{Q_{\sz}(\tau)}$. [Note that considering the
average quantity eliminates the pathological behavior of individual
atypical circuits -- such as for example one comprised of only
identity gates -- because their contribution are only included in the
average with vanishingly small probability.] Moreover, the average of
the string density over circuits depends on the initial condition,
${\beta^{\rm z}}$, but is independent of the site index $i$. Also,
note that we kept the subscript ${\sz}$ on ${Q_{\sz}(\tau)}$ as a
reminder of the initial ${\sz}$-string expectation value in
Eq.~\eqref{eq:Qzi}. Parametrizing the $\tau$ dependence in terms of
the number of layers of gates $\ell$ (the depth) of the circuit
describing the unitary transformation $\hat U$ that generated the
evolution of string amplitudes up to time $-\tau$, we can define
\begin{align}
\label{eq:Eps}
  \epsilon(\ell;\beta^{\rm z}) \equiv {1-\frac{4}{3} \rho(\ell; \beta^{\rm z})}
   \;,
\end{align}
to write $\overline{Q_{\sz}(\tau)}= \frac{1}{2^n}\sum_{\beta^{\rm z}}
\epsilon(\ell;\beta^{\rm z})$. A lower bound on the function $\eta(n)$
in Eq.~\eqref{eq:pseudorandom-condition} is determined by how fast
$\rho (\ell; \beta^{\rm z})$ reaches its asymptotic value of $3/4$
starting from an initial condition associated with the arbitrary
(non-trivial) initial string state $\beta^{\rm z}$.

The equation describing the evolution of the average string weight
$\rho (\ell; \beta^{\rm z})$ can be derived by following all
$15\ (=4\times 4 -1)$ non-trivial two-site strings through the unitary
evolution with consecutive layers of effective (average) gates which
connect with equal amplitude each of these states to themselves and to
each other. Here we shall make a mean-field approximation, which is
equivalent to the assumption that the densities at different positions
along the string are uncorrelated. It then follows that, since the
identity string does not scatter into a non-trivial string, a
configuration involving identity operators on both sites, which occurs
with probability $(1-\rho)^2$, cannot contribute a Pauli operator on a
given site. Otherwise, with probability $1-(1-\rho)^2$, non-trivial
$1$-site and $2$-site string states scatter into a configuration with
a Pauli operator on a given site with transition probability
$12/15=4/5$, accounting for the fact that only $12$ ($3$ weight-1
strings and $9$ weight-2 strings) out of the $15$ non-trivial string
states feature a Pauli operator on that site. Therefore,
\begin{align}
  \label{eq:MF-2qubit}
  \rho(\ell+1; \beta ^{\rm z})
  =&\;\;\;\;
     \left(1-\rho(\ell;\beta ^{\rm z})\right)^2 \times 0
     \nonumber\\
  &+
     [1-\left(1-\rho(\ell; \beta ^{\rm z})\right)^2]\times\frac{4}{5}
     \nonumber\\
  \equiv&\;\;\;\;
     \frac{4}{5}\;\rho(\ell; \beta^{\rm z})\;(2-\rho(\ell;\beta ^{\rm z}))
     \;.
\end{align}
As expected, $\rho (\ell \rightarrow \infty;\beta^{\rm z}) =3/4$ is a
fixed point. Writing Eq. (\ref{eq:MF-2qubit}) in terms
  of $\epsilon(\ell;\beta ^{\rm z})$ defined in Eq.~\eqref{eq:Eps},
we obtain
\begin{align}
  \label{eq:MF-epsilon}
  \epsilon(\ell+1;\beta ^{\rm z})
  =&
     \frac{2}{5}\;\epsilon(\ell;\beta ^{\rm z})+\frac{3}{5}\;(\epsilon(\ell;\beta ^{\rm z}))^2
     \;.
\end{align}
This equation must be solved with initial condition
$\epsilon(0;\beta ^{\rm z})=1-(4/3)\,\rho (0;\beta^{\rm
  z})$. Asymptotically, $\epsilon(\ell;\beta ^{\rm z})$ tends to zero
exponentially in $\ell$ as $\sim(2/5)^\ell$, for any initial string
state $\beta ^{\rm z}$. 

We note that incorporating two-site correlations beyond mean-field can
alter the coefficients in Eq.~\eqref{eq:MF-epsilon} by terms of order
$1/n$ but cannot change the fixed point density,
$\rho (\ell \rightarrow \infty;\beta^{\rm z}) =3/4$. In particular,
these $1/n$ corrections can modify the coefficient of the linear term
in Eq.~ \eqref{eq:MF-epsilon} but cannot eliminate it all together,
thus preserving the exponential decay of OTOCs with $\ell$. For
$\ell\sim{\cal O}(\log n)$, $\overline{Q_{\sz}}$ in
Eq.~\eqref{eq:Qaverage} can only decay as a power law in $n$,
violating the assumption that $\eta(n)$ in
Eq.~\eqref{eq:pseudorandom-condition} is superpolynomially small in
$n$, as pseudorandomness requires. We conclude that evolution via a
$\log n$-depth circuit built from universal 2-qubit gates drawn
uniformly from unitaries in $U(4)$ is incapable of reaching a
pseudorandom state. [It is interesting to note that the
  number of layers required to reach the equilibrium string weight
  $\rho = {3/4}\,(1 - \epsilon)$ starting from an initial value
  $\rho (0;\beta ^{\rm z} = \sz_i) \sim {1/n}$ that emerges from the
  differential equation derived from the mean-field recursion in
  Eq.~\eqref{eq:MF-2qubit} corresponds to a circuit of size
  $S = \frac{5}{6}\, n \left[\ln n + \ln (1/{\epsilon}) \right]$. This
  is precisely the expression for the lower bound on circuit size
  required for anti-concentration derived in Ref.~\cite{Brandao2022}
  for 2-qubit gates on a complete graph and conjectured earlier by
  Harrow and Mehraban~\cite{Harrow2018approximate}.]

In the next section we use the ${\cal O}(\log n)$-depth 
structured classical reversible circuits discussed in
Ref.~\cite{cipher-paper} to build a pseudorandom quantum state, i.e.,
a quantum state for which the bound in
Eq.~\eqref{eq:pseudorandom-condition} is satisfied.


\vspace{.2cm}
\noindent{\bf{Structured random quantum circuits:}}
\label{sec:structured} Let us start with a product state $\ket{0}^{\otimes n}$ in the
computational basis, and apply a non-trivial string $\beta^{\rm x}$ of Pauli $\sx$
operators that flips the initial state to $\ket{\beta^{\rm x}}$. By
applying Hadamard gates to this state we then obtain:
\begin{align}
  H^{\otimes n}\;\ket{\beta^{\rm x}} = \frac{1}{\sqrt{2^{n}}}\sum_x\;(-1)^{\beta^{\rm x}\cdot x}\;\ket{x}
  \;.
\end{align}
Finally, evolving the resulting state with a classical reversible permutation circuit
$\hat P$, $\hat P\,\ket{x}=\ket{P(x)}$, leads to:
\begin{align}
  \label{eq:prwf}
  \ket{\psi_{\beta^{\rm x}}}
  &=
    \frac{1}{\sqrt{2^{n}}}\sum_x\;(-1)^{\beta^{\rm x}\cdot x}\;\ket{P(x)}
    \nonumber\\
  &=
    \frac{1}{\sqrt{2^{n}}}\sum_x\;(-1)^{\beta^{\rm x}\cdot P^{-1}(x)}\;\ket{x}  
  \;.
\end{align}
A general reversible
classical circuit $\hat P$ can be built from 3-bit gates in $S_8$,
which generate all permutations on the space of $n$ bits within the
alternating group $A_{2^n}$ (all even permutations in the group
$S_{2^n}$).

Refs.~\cite{Ji-Liu-Song2018,Brakerski-Shmueli2020} show that a state
of the form in Eq.~\eqref{eq:prwf}, with the phase given by a
pseudorandom function, is a pseudorandom state. Here we will use
pseudorandom permutations that, for large $n$, cannot be distinguished
from pseudorandom functions.  More precisely, we will deploy 3-stage
$\log n$-depth circuits discussed in Ref.~\cite{cipher-paper}, where
it was argued that such circuits generate permutations satisfying the
necessary (and conjectured to also be sufficient) conditions for
pseudorandomness. The resulting quantum circuit architecture that
generates the pseudorandom quantum states considered in this section
is shown in Fig.~\ref{fig:pseudorandom}.

To illustrate the importance of the $3$-stage structure to the
generation of pseudorandomness, we consider the expectation value of a
Pauli string operator in the state $\ket{\psi_{\beta^{\rm x}}}$ of
Eq.~\eqref{eq:prwf}. We proceed by applying $\S_\alpha$ to this state:
\begin{widetext}
\begin{align}
  \S_\alpha\;\ket{\psi_{\beta^{\rm x}}}
  &=
    \frac{1}{\sqrt{2^{n}}}\sum_x\;(-1)^{\beta^{\rm x}\cdot P^{-1}(x)}\;\S_\alpha\ket{x}
    \nonumber\\
    &=
      \frac{1}{\sqrt{2^{n}}}\sum_x\;(-1)^{\beta^{\rm x}\cdot P^{-1}(x)}\;
      (-1)^{\alpha^{\rm z}\cdot x}\ket{x\oplus \alpha^{\rm x}}
    \nonumber\\
    &=
      (-1)^{\alpha^{\rm z}\cdot \alpha^{\rm x}}\frac{1}{\sqrt{2^{n}}}\sum_x\;(-1)^{\beta^{\rm x}\cdot P^{-1}(x\oplus \alpha^{\rm x})}\;
      (-1)^{\alpha^{\rm z}\cdot x}\ket{x}
      \;, 
\end{align}
which, in turn, leads to the following expression for the expectation
value of the string operator $\S_\alpha$:
\begin{align}
  \bra{\psi_{\beta^{\rm x}}}\;\S_\alpha\;\ket{\psi_{\beta^{\rm x}}}
  &=
    (-1)^{\alpha^{\rm z}\cdot \alpha^{\rm x}}
    \frac{1}{2^{n}}\sum_x\;(-1)^{\beta^{\rm x}\cdot [P^{-1}(x)\oplus P^{-1}(x\oplus \alpha^{\rm x})]}\;
      (-1)^{\alpha^{\rm z}\cdot x}
      \;.
\end{align}
We note that for $\alpha^{\rm z}=0$, $\alpha^{\rm x}_k = \delta_{k,i}$
(i.e., $\S_{\alpha}=\sx_i$), and $\beta^{\rm x}_l = \delta_{l,j}$
(i.e., flipping only the $j$th qubit of the initial state) this
expectation value is expressed as the OTOC representing the
SAC~\cite{cipher-paper}, a simple test of security for a classical
block cipher:
\begin{align}
\label{eq:S-OTOC}
  Q^{\rm SAC}_{ij}
  &\equiv
    \frac{1}{2^{n}}\sum_x\;(-1)^{[P_j^{-1}(x)\oplus P_j^{-1}(x\oplus c_i)]}\;
    \;,
\end{align}
where the qubit-wise XOR operation for two $n$-qubit strings
$x\oplus c_i$ flips the $i$th qubit of $x$ (i.e., $c_i=2^i$).
\end{widetext}

In Ref.~\cite{cipher-paper} we presented a calculation of the
evolution of the SAC OTOC Eq.~\eqref{eq:S-OTOC} through the
application of consecutive layers of the structured cipher.  To
summarize the results of that calculation, we first introduce $\hat
P^{-1}(\ell)$, the partial circuit comprised of the first $\ell$
layers of the circuit $\hat P^{-1}$, with $\hat P^{-1}(0)\equiv
\openone$ and $\hat P^{-1}(\ell_f)\equiv \hat P^{-1}$ and define the
expectation value of the string operator in Eq.~\eqref{eq:S-OTOC}
after $\ell$ layers of the permutation $P^{-1}$ are applied as,
\begin{align}
  Q^{\rm SAC}_{ij}(\ell)
  &=
    \frac{1}{2^{n}}\sum_x\;(-1)^{[P_j^{-1}(x,\ell)\oplus P_j^{-1}(x\oplus c_i,\ell)]}
    \;.
\end{align}
As in the mean-field calculation above we will focus on averages over
circuits, $s(\ell)=\overline{Q^{\rm SAC}_{ij}(\ell)}$ and
$q(\ell)=\overline{[Q^{\rm SAC}_{ij}(\ell)]^2}$. Since gates defining
individual layers are chosen independently we can easily derive
recursion relations relating $s(\ell+1)$ and $q(\ell+1)$ to $s(\ell)$
and $q(\ell)$, which depend on the specific gate set chosen. As shown
in Ref.~\cite{cipher-paper} and summarized in
the `SAC OTOC' discussion of the Methods section, evolution through $\ell$ layers
of linear inflationary gates, leads to
\begin{subequations}
\begin{align}
  \label{eq:s-recursion}
  s(\ell+1)
  = \frac{2}{3}\,[s(\ell)]^2 + \frac{1}{3}\,[s(\ell)]^3
  \;,
\end{align}
\begin{align}
  q(\ell+1)
  =
  \frac{2}{3}\,[q(\ell)]^2 + \frac{1}{3}\,[q(\ell)]^3
  \;;
  \label{eq:recursion_inflationary_q}
\end{align}
\end{subequations}
and evolution through $\ell$ layers of supernonlinear
gates, which maximize string entropy production`\cite{cipher-paper}, leads to:

\begin{subequations}
\begin{align}
\label{eq:recursion s2}
  s(\ell+1)
  =
  \frac{3}{7}\,s(\ell)
  +
  \frac{3}{7}\,[s(\ell)]^2
  +
  \frac{1}{7}\,[s(\ell)]^3
  \;,
\end{align}

\begin{align}
 \label{eq:recursion_supernonlinear_q}
  q(\ell+1)
  =&\;
  \frac{3}{28}\,
     \left([s(\ell)]^2 + [s(\ell)]^3\right)\\
   &+ \frac{3}{28}\,q(\ell) \,
     \left(1+ 2\,s(\ell) + 2\,[s(\ell)]^2\right)\nonumber\\
   &+ \frac{3}{28}\,[q(\ell)]^2\,
     \left(1 + s(\ell)\right)
     + \frac{1}{28}\,[q(\ell)]^3
  \;.\nonumber
\end{align}
\end{subequations}
We note that the recursion relations in Eqs.~\eqref{eq:s-recursion}, ~\eqref{eq:recursion_inflationary_q}, ~\eqref{eq:recursion s2}, and \eqref{eq:recursion_supernonlinear_q} are exact for tree-structured circuits of depth $\ell \le \log _3 n$.

We note that if the circuit contained only supernonlinear gates, the
analysis of the decay of the OTOC (and of the expectation value of the
string operator) could be carried out by linearizing
Eqs.~\eqref{eq:recursion_supernonlinear_q} for small $s$ and $q$:
\begin{subequations}
\begin{align}
  s(\ell+1)
  =
  \frac{3}{7}\,s(\ell)
  +\cdots
  \;,
\end{align}
\begin{align}
  q(\ell+1)
  =&\frac{3}{28}\,q(\ell) +\cdots
  \;.
\end{align}
\end{subequations}
In this case, it is inescapable that $q(\ell)$ can only decay
exponentially with depth $\ell$: $q(\ell)\sim e^{-\lambda\ell}$, with
$\lambda = \ln (28/3)$. One can interpret the coefficient of the
linear term in the expansion of the recursion relation as a Lyapunov
exponent, $\lambda$. The exponential decay of $q(\ell)$ with a finite
Lyapunov exponent $\lambda$ implies that circuits of $\log n$ depth
can only lead to polynomial decay of the SAC OTOC. It is important to
stress that the same linear leading behavior in $s$ and $q$ of the
recursion relations occurs for random
circuits of universal gates. Eliminating the linear terms requires fine tuning - this is
precisely what makes the linear inflationary gates both special and
necessary for ensuring that the SAC OTOC decays exponentially with $n$
for depth $\log n$ structured circuits.

Indeed, the recursions for $s$ and $q$ in the case of inflationary
gates start with quadratic leading terms. [Note that $q=1$ is a fixed
point of the recursion Eq.~\eqref{eq:recursion_inflationary_q}, and
thus nonlinear gates are needed to reduce the value of $q$ below 1
before the system can evolve towards the $q=0$ fixed point.] To lowest
order in $q$, the recursion Eq.~\eqref{eq:recursion_inflationary_q},
which is activated following the action of the layers of
supernonlinear gates, reads
\begin{align}
  q(\ell+1)
  =
  \frac{2}{3}\,[q(\ell)]^2 + \cdots
  \;,
\end{align}
the asymptotic solution of which is a double exponential in $\ell$,
$q(\ell)\sim \frac{3}{2}\left[\frac{2}{3}\;q(0)\right]^{2^\ell}$. This
behavior, which corresponds to an infinite Lyapunov exponent, is
non-universal but essential in ensuring the exponential decay with $n$
of the SAC OTOC and, equivalently, in proving the superpolynomial bound
of Eq.~\eqref{eq:pseudorandom-condition} for expectation values of
string operators in the quantum state $\ket{\psi_{\beta^{\rm x}}}$ in
Eq.~\eqref{eq:prwf}.

We note that there are 144 3-bit inflationary gates among the $8!$
gates $3$-bit gates in $S_8$~\cite{cipher-paper}. One can then ask
whether inflationary gates are also present among the $2$-qubit $U(4)$
gates that generate universal quantum computation, in which case one
could imagine constructing pseudorandom quantum states by employing
such 2-qubit gates. Below we show that there are no 2-qubit
inflationary gates, but that $2$-qudit
inflationary gates do exist for $d\ge 3$ and $d$ prime.


\subsection*{Absence of two-qubit inflationary gates}
\label{sec:theorem}

The main message of this section is that there are no inflationary
$2$-qubit gates in $U(4)$, i.e., that there are no $U(4)$ gates which
eliminate the stay probability of weight-$1$ strings. We prove this
statement first for $2$-qubit Clifford gates, and then for general
unitary gates in
$U(4)$.  \\

\noindent{\bf{Clifford Gates:}} We argue by contradiction: suppose that a
two-qubit Clifford gate $U_{\rm Cl}$ maps both Pauli operators
$\sx_1$ and $\sz_1$ on site $1$ to Pauli strings of weight
two with a footprint on both site 1 and site 2:

\begin{eqnarray}
  U^\dagger_{\rm Cl} \;\sx_1 \;U_{\rm Cl} &= \hat\sigma_1^\alpha\; \hat\sigma_2^\beta\;,
                                                 \nonumber\\ 
  U^\dagger_{\rm Cl} \;\sz_1 \;U_{\rm Cl} &= \hat\sigma_1^\mu \; \hat\sigma_2^\nu\;.
\end{eqnarray}
Since the anticommutation relation is preserved under gate
conjugation, $\{ \hat\sigma_1^\alpha \hat\sigma_2^\beta,
\hat\sigma_1^\mu \hat\sigma_2^\nu\}=0$, the operator content of the
two Pauli strings must be identical on one site, i.e., one must have
either $\alpha=\mu, \beta \neq \nu$ or $\beta = \nu, \alpha \neq
\mu$. By considering the transformation of the commutator it
immediately follows that $U_{\rm Cl}$ maps $\hat\sigma_1^{\rm y}$ to a
single-site Pauli operator, thus contradicting the initial assumption
that $U_{\rm Cl}$ maps all weight-1 strings to weight-2 strings.  \\

\noindent{\bf{$\bf{U(4)}$ Unitaries:}} We start with a special case which we
then use to establish the general result. Again, we argue by
contradiction: we assume that a 2-qubit unitary maps $\sx_1$ into a
single weight-$2$ Pauli string, $\hat{\mathcal{S}}$, and maps $\sz_1$
to a superposition of weight-$2$ Pauli strings:
\begin{eqnarray}
U^\dagger \;\sx_1 \;U &=& \hat{\mathcal{S}}  \nonumber \\
  U^\dagger \;\sz_1 \;U &=& \sum_\mu \;b_\mu^{c} \;\hat{\mathcal{S}}^\mu_c + \sum_\nu \;b_\nu^{a} \;\hat{\mathcal{S}}^\nu_{a},
\label{eq:22}
\end{eqnarray}
where, in the second equation, the $b_\mu^{c}$ and $b_\nu^{a}$ are
string amplitudes, associated separately with Pauli strings that
commute or anticommute with $\hat{\mathcal{S}}$: $[\hat{\mathcal{S}},
  \hat{\mathcal{S}}^\mu_c]=0$ and $\{\hat{\mathcal{S}},
\hat{\mathcal{S}}^\nu_a \}=0$. Again, the conjugated Pauli operators
must satisfy: $\{U^\dagger \,\sx_1 \,U\,,\, U^\dagger\, \sz_1 \,U
\}=0$.  This condition is automatically satisfied by
$\hat{\mathcal{S}}^\nu_a$, whereas for $\hat{\mathcal{S}}^\mu_c$ it
requires
\begin{equation}
2 \,\sum_\mu \,b_\mu^c \,\hat{\mathcal{S}} \,\hat{\mathcal{S}}^\mu_c = 0.
\label{eq:23}
\end{equation}
We next consider the evolution of $\hat\sigma_1^{\rm y}$:
\begin{equation}
U^\dagger\, \hat\sigma_1^{\rm y} \,U \propto \sum_\mu \,b_\mu^{c}\,  \hat{\mathcal{S}} \,\hat{\mathcal{S}}^\mu_c + \sum_\nu \,b_\nu^{a} \,\hat{\mathcal{S}} \,\hat{\mathcal{S}}^\nu_{a}.
\label{eq:24}
\end{equation}
Notice that the summation in the second term of Eq.~\eqref{eq:24} must
involve operators of weight 1 for the same reason as explained above:
the operator content of $\hat{\mathcal{S}}$ and
$\hat{\mathcal{S}}_a^\nu$ must be identical on one site in order for
these operators to anticommute, whereas the first term must vanish
according to Eq.~\eqref{eq:23}. Hence, we reach a contradiction,
namely that the inflationary condition of Eqs.~\eqref{eq:22} cannot be
satisfied for all Pauli operators (weight-$1$ Pauli strings.)

Finally, we consider the general case in which the unitary
transformation evolves $\sx_1$ into a superposition of weight-$2$
strings:
$U^\dagger \,\sx_1 \,U = \sum_{\alpha\beta}\,M_{\alpha \beta} \;\hat\sigma_1^\alpha\,
\hat\sigma_2^\beta$, where $M$ is a $3\times 3$ real matrix. One can
perform a singular value decomposition, $M = A \,\Lambda \,B^\top$, which
amounts to a basis rotation of the single-site Pauli operators:
$\hat{\tilde{\sigma}}_1^{\beta} = \sum_{\alpha}\,\hat\sigma_1^{\alpha}\;A_{\alpha \beta}$,
$\hat{\tilde{\sigma}}_2^{\beta} = \sum_{\alpha}\,\hat\sigma_2^{\alpha} \;B_{\alpha\beta}$.
In the new basis, the Pauli strings are diagonal:
\begin{equation}
  U^\dagger \;\sx_1 \;U
  =
  \lambda_x \;\hat{\tilde{\sigma}}_1^{\rm x} \,\hat{\tilde{\sigma}}_2^{\rm x}
  +
  \lambda_y \;\hat{\tilde{\sigma}}_1^{\rm y} \,\hat{\tilde{\sigma}}_2^{\rm y}
  +
  \lambda_z \;\hat{\tilde{\sigma}}_1^{\rm z} \,\hat{\tilde{\sigma}}_2^{\rm z}
  \;.
  \label{eq:26}
\end{equation}
However, since the right hand side of Eq.~\eqref{eq:26} must square to
the identity, two of the $\lambda$'s must be zero while the other one
must be equal to unity. Thus, we have reduced the general case to the
special case where $\sx_1$ evolves into a single weight-$2$ string, as
in Eq.~\eqref{eq:22}. Thus, we proved the main assertion of this
section, namely that if one restricts oneself to 2-qubit unitary
gates, there will always be a finite stay probability for weight-$1$
strings. As already discussed, in turn, this prevents one from
reaching pseudorandom quantum states with $\log n$-depth circuits.


\subsection*{Existence of two-qudit inflationary gates for $q\ge 3$}
\label{sec:inflationary-quantum}

An important conclusion of this paper, which suggests a circuit design
for realizing fast and thorough quantum scramblers, is that it is
always possible to construct inflationary 2-qudit Clifford unitaries
which transform all single-site generalized Pauli operators
(weight-$1$ generalized Pauli strings) into weight-$2$ generalized
Pauli strings. The proof of this result is given in
the `Two-qudit Inflationary Clifford Gates' discussion of the Methods section for a subset of unitaries in
$U(q^2)$ for which the local Hilbert-space dimension $d\geq3$ and $d$
prime.

\begin{figure}[ht]
\centering
\includegraphics[width=0.4\textwidth]{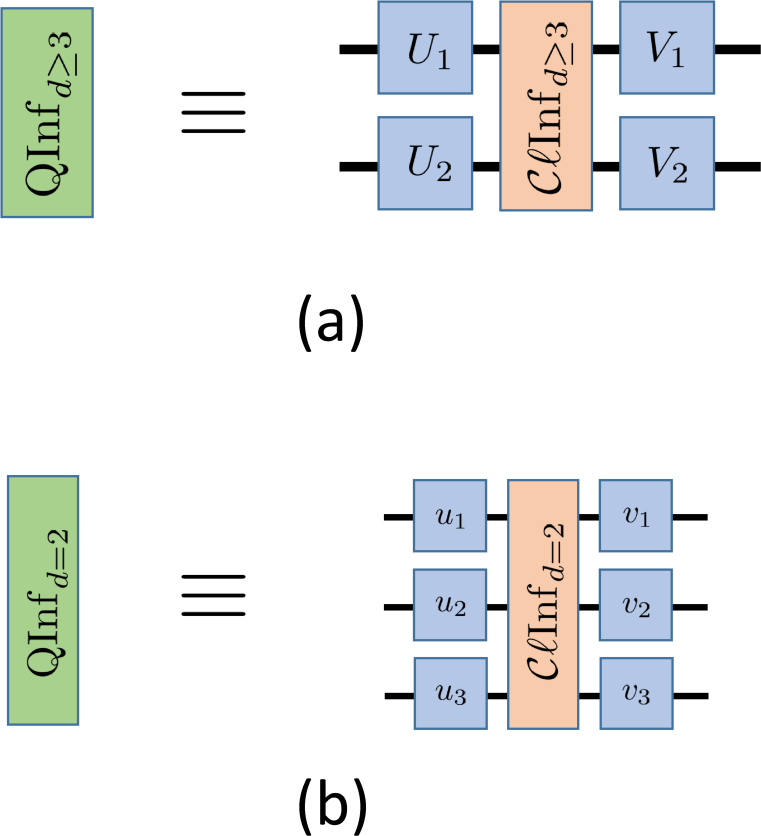}
\caption{{\bf Inflationary Quantum (IQ) Gates:} (a) 2-qudit IQ gates
  obtained by padding Clifford qudit inflationary gates (see
  the second section of Methods) that transform weight-1 strings
  into weight-2 strings gates with 1-qudit rotations at inputs and
  outputs; (b) 3-qubit IQ gates obtained by adopting the 144 3-bit
  classical (linear) inflationary gates that transform weight-1
  strings into weight-2 strings (see Ref.~\cite{cipher-paper}) to
  qubits, and padding the resulting 3-qubit gates with 1-qubit
  rotations at inputs and outputs. Employing random circuits of
  3-qubit or 2-qudit IQ gates will eliminate the stay probability of
  weight-1 Pauli strings, as depicted in the inset to
  Fig.~\ref{fig:stay_prob}, while, at the same time proliferating
  operator strings and generating string entropy.}
\label{fig:inflationary-gate}
\end{figure}

Padding such a 2-qudit Clifford gate with 1-qudit rotations at inputs
and outputs, as depicted in Fig.~\ref{fig:inflationary-gate}, leads to
special inflationary quantum (IQ) gates, to which we already referred
in the introduction and in the `Our Contributions' section above. Because of
their inflationary property, the 2-qudit Clifford gates discussed discussed in this section are necessarily entangling. These
2-qudit Clifford gates also form a finite group, which includes the
identity gate, and therefore one can write single qudit unitaries as
products of IQ gates.  As demonstrated in
Ref.~\cite{brylinski2001universal}, an entangling 2-qudit gate and
arbitrary single qudit rotations are the two ingredients required for
the universality of a gate set. Therefore, the 2-qudit IQ gates in
Fig.~\ref{fig:inflationary-gate} form a universal set for quantum
computation.

IQ gates can also be realized by padding the 144 classical
inflationary gates of Ref.~\cite{cipher-paper} (shown in Fig.~\ref{fig:CNOT-equiv-inflationary})
with 1-qubit rotations at inputs and
outputs, as depicted in Fig.~\ref{fig:inflationary-gate}b. We note
that the 144 classical inflationary gates generate all classical
linear 3-bit gates and, in particular, the identity gate and 2-bit
CNOTs across any of the 3 bitlines, [The inflationary gates
  associated to the two permutations 0~3~5~6~7~4~2~1 and
  1~4~6~3~2~7~5~0 suffice to generate the group of all 1344
  permutations associated to classical linear 3-bit gates, i.e., gates
  $g$ such that $g(x\oplus y)= g(x)\oplus g(y)\oplus c$, for a
  constant $c$.] One can then write both single qubit unitaries and
entangling CNOTs as products of IQ gates and thus, the 3-qubit IQ
gates in Fig.~\ref{fig:inflationary-gate}b also form a universal set
for quantum computation.

As discussed in the `Our contributions' section above and in more detail in
Ref.~\cite{cipher-paper}, reaching cryptographic-level scrambling with
$\log n$-depth classical reversible circuits required a 3-stage
structure that separated linear classical gates responsible for string
inflation from nonlinear classical gates responsible for string
proliferation and entropy production.
What makes these 2-qudit and 3-qubit IQ gates special is that they
generate both `diffusion' and `confusion' in the sense of
Shannon~\cite{Shannon1949}, i.e., IQ gates posses the ability to
simultaneously (a) eliminate stay-probabilities for weight-1 strings
and accelerate the inflation of strings; and (b) proliferate the
number of strings and generate string entropy. We thus expect that
single-stage random quantum circuits comprised of IQ gates can
scramble both at the speed limit (i.e., in $\log n$-depth) and to
cryptographic level.


\section*{Discussion}
\label{sec:conclusions}

As we detailed in this paper, scrambling rapidly (i.e., with
$\log n$-depth) and thoroughly (i.e., to cryptographic precision), via
quantum circuits, is a tall order. More precisely, this paper makes
two specific complementary points, namely: (i) that generic
$2$-qubit-gate quantum circuits cannot scramble information to
cryptographic precision within a computational time scaling as
$\log n$; and (ii) that fast scrambling to cryptographic precision can
be realized with a special set of universal inflationary quantum (IQ)
gates. These special IQ gates can simultaneously expand
individual Pauli strings as well as proliferate their number, the
latter leading to string entropy production.

\begin{figure}[!t]
\centering
\includegraphics[width=0.4\textwidth]{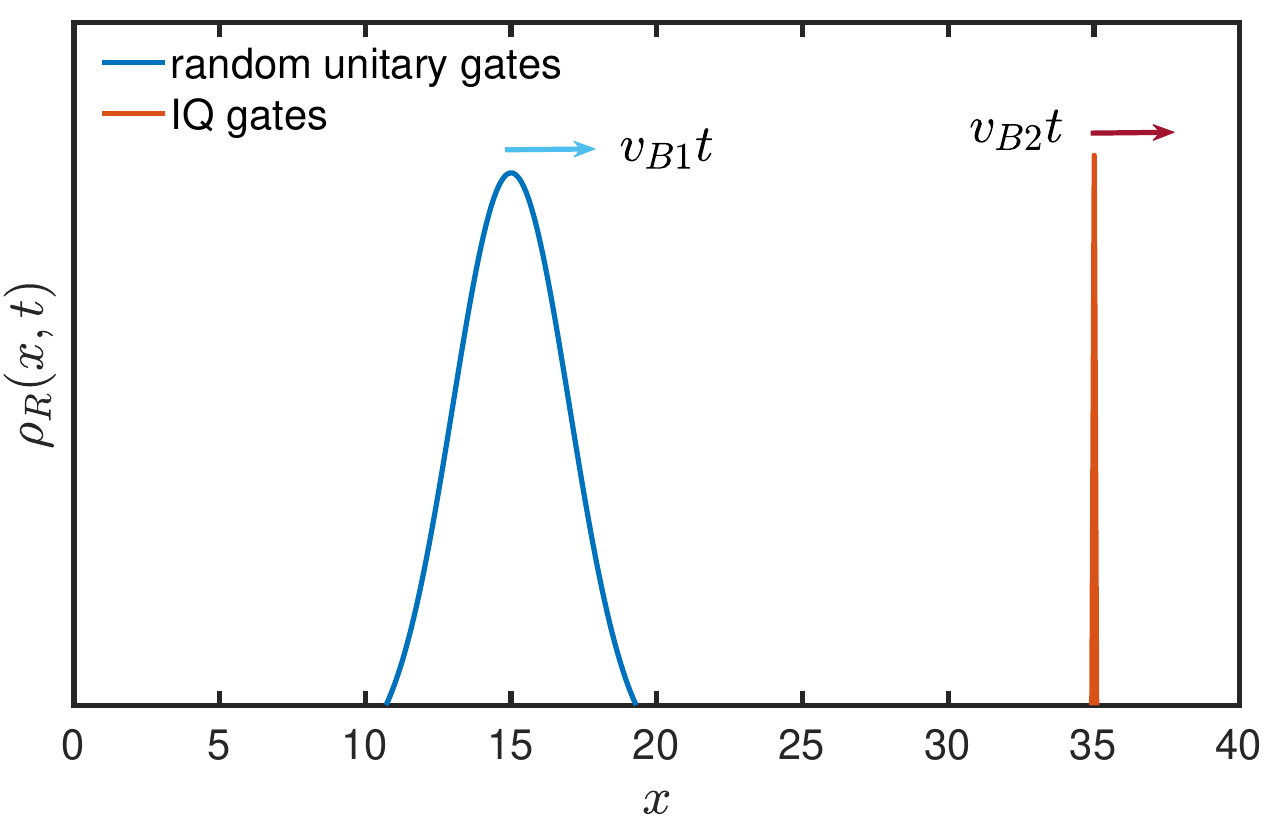}
\caption{{\bf Operator front profile evolution.} Illustration of the profile of the right operator front $\rho_R(x,t)$ (i.e.  total weight of Pauli strings with right endpoint at $x$) under random unitary circuit and IQ circuit evolution.  IQ circuits lead to a larger butterfly velocity and an absence of operator front broadening compared to random unitary circuits. }
\label{fig:fronts}
\end{figure}

IQ gates should play a key role in a number of areas in quantum
information in which fast scrambling is desirable. (We note that the
special properties of IQ gates will affect the behavior of circuits in
any architecture, beyond the tree-structured circuits of long-ranged
gates on which we concentrated in this paper.)  For example, in the
case of one-dimensional `brickwall' circuits of 2-qudit IQ gates,
the front associated with operator spreading will propagate 
deterministically without dispersion, at the Lieb-Robinson speed
limit. This behavior is due to the fact that at the front (i.e., at
the edges of the Pauli strings at the light cone boundary) an IQ gate
always acts on a fresh site with vanishing string weight (i.e., an up
to then untouched site just outside the front), so that the
inflationary property ensures that the evolved string will always
acquire weight at that site after evolution by the gate. By contrast,
as described in Ref.~\cite{Nahum-etal2017}, evolution by generic qudit
circuits, would lead to a stochastic evolution of the front, with an
average velocity below the maximum attainable value, and with a
front-width that spreads diffusively. A cartoon of the difference
between these cases is shown in Fig.~\ref{fig:fronts}. More generally,
IQ gates would lead to faster, deterministic front propagation in any
spatial dimension $D$. The speed up is most dramatic when
$D\to \infty$ as in the case of our tree-structured circuits, for
which IQ gates are essential for reaching cryptographic-level
scrambling with minimal $\log n$-depth circuits.

Furthermore, we expect that the rapid scrambling property of IQ gates
provides an additional ingredient that should lead to stronger bounds on
$t$-designs. For example, a circuit of IQ gates may validate the
conjecture of Harrow and Mehraban~\cite{Harrow2018approximate} that
one can build 2-designs with $\log n$-depth circuits.

IQ gates may also be useful in desiging novel quantum advantage
experiments, since they accelerate the expansion and proliferation of
Pauli strings. We note, however, that inflation of strings are
counter-acted by depolarizing noise, which removes contributions from
high weight strings. This mechanism of suppression of large strings
has been explored in Refs.~\cite{Aharonov-etal,Gao-etal} to design
efficient classical algorithms for sampling from the output
distribution of a noisy random quantum circuit.

Finally, while employing random circuits of IQ gates should enable the
construction of cryptographic level fast quantum scramblers -
quantum `superscramblers' - we do not expect that unitary evolution
via a time-independent Hamiltonian of interacting qudits or qubits can
scramble to such a level in a time ${\cal O}(\log n)$, even if
non-local couplings are employed.

\section*{Methods}




\subsection*{Tree-structured circuits}
\label{sec:appendix-tree-structure}

Here we present the wiring of tree-structured circuits that both
accelerate the scrambling and allowed us to obtain analytically the recursion relations (18a,18b,19a,19b,20a,20b,21), 
the detailed derivation of which we present below.] Tree-structure
circuits connect qubits or, more generally, qudits in a hierarchy of
scales, and mimic systems in $D\to \infty$ spatial dimensions.

We consider first a tree-structured circuit in which pairs of qudit
indices acted by 2-qudit gates are arranged in a hierarchical (tree)
structure. Let us consider the case when the number of qudits, $n$, is
a power of 2, $n=2^q$. Each level in the tree hierarchy comprises of a
layer with $n/2$ 2-qudit gates. We proceed by forming pairs indices
for each layer $\ell$ of gates, selected as follows:
\begin{align}
  \ell = 1: & \quad(0,1)\;(2,3)\;(4,5)\;(6,7) \dots \nonumber\\
  \ell = 2: & \quad(0,2)\;(1,3)\;(4,6)\;(5,7) \dots \nonumber\\
  \ell = 3: & \quad(0,4)\;(1,5)\;(2,6)\;(3,7) \dots \nonumber\\
  \ell = 4: & \quad(0,8)\;(1,9)\;(2,10)\;(4,11) \dots \nonumber\\
  \dots \quad&
\end{align}
More precisely, each of the $n/2=2^{q-1}$ pairs in layer $\ell$ are
indexed by $(i,j)$, which we write in base 2 as
\begin{align}
  i=&z_0 + 2\;z_1+2^2\;z_2+\dots+2^{\ell-1}\times\underline{0}+\dots\;2^{q-1}\;z_{q-1}\nonumber\\
  j=&z_0 + 2\;z_1+2^2\;z_2+\dots+2^{\ell-1}\times\underline{1}+\dots\;2^{q-1}\;z_{q-1}
  \;,
      \label{eq:trit-triples}
\end{align}
where $z_a=0,1$, for $a=0,\dots, q-1$. Notice that at layer $\ell$ the
members of the pairs, $(i,j)$, are numbers that only differ in the
$(\ell-1)$-th bit, while the other $q-1$ bits $z_a, a\ne \ell-1$,
enumerate the $2^{q-1}=n/2$ pairs. (If more than $q$ layers are
needed, we recycle in layer $\ell > q$ the pairs of layer
$\ell\!\!\mod q$.)

\begin{figure}[!t]
\centering
\includegraphics[width=0.45\textwidth]{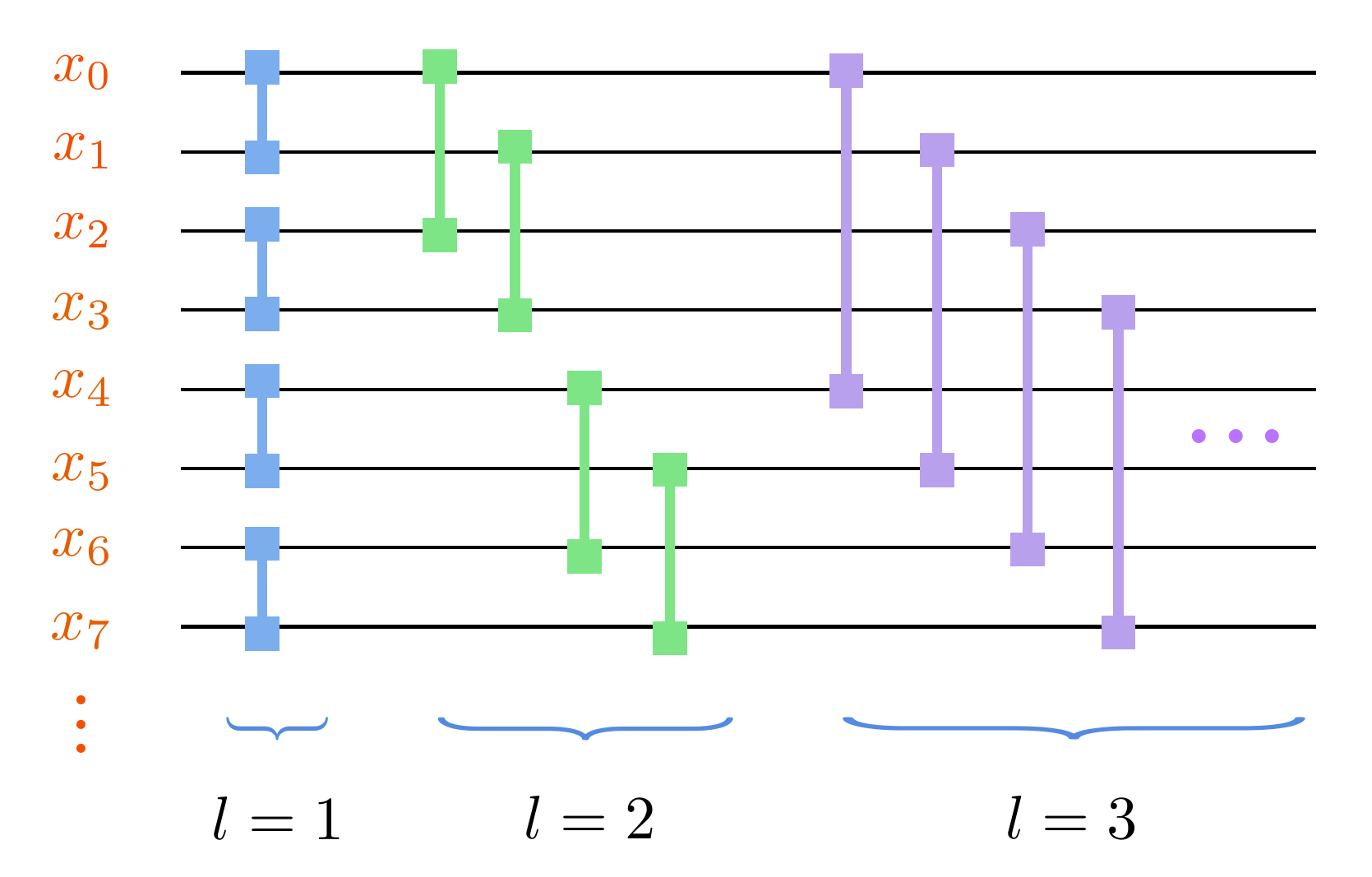}
\caption{{\bf Hierarchical tree-structured circuit consisting of two-qudit unitary gates.} Each unitary gate is represented as a solid line with square endpoints, which indicate the two qudits on which the gate acts. Each layer contains $n/2$ gates acting on different non-overlapping pairs of qudits. Circuits consisting of three-qubit gates that form a ternary tree structure, like the circuits $\hat{L}_{l/r}$ and $\hat{N}$ shown in Fig. 2, can be constructed in a similar fashion. }
\label{fig:circuit_tree}
\end{figure}

Once the pairs of indices, $(i,j)$, are selected for each layer, we
can generate other similar binary trees by mapping $(i,j)$ onto
$\left(\pi(i),\pi(j)\right)$, via a (randomly chosen) permutation
$\pi$ of the $n$ indices. A schematic of the hierarchical tree-structure presented above is shown in 
Fig.~\ref{fig:circuit_tree} below.


The construction can be generalized to trees of other degrees, for
example the ternary tree introduced in Ref.~\cite{cipher-paper}, which
we borrow to provide an additional example. Consider the case when $n$
is a power of 3, $n=3^q$. In the ternary case, we proceed by forming
groups of triplets of indices for each layer, selected as follows:
\begin{align}
  \ell = 1: & \quad(0,1,2)\;(3,4,5)\;(6,7,8) \dots \nonumber\\
  \ell = 2: & \quad(0,3,6)\;(1,4,7)\;(2,5,8) \dots \nonumber\\
  \ell = 3: & \quad(0,9,18)\;(1,10,19)\;(2,11,20) \dots \nonumber\\
  \ell = 4: & \quad(0,27,54)\;(1,28,55)\;(2,29,56) \dots \nonumber\\
  \dots \quad&
\end{align}
More precisely, each of the $n/3=3^{q-1}$ triplets in layer $\ell$ are
indexed by $(i,j,k)$, which we write in base 3 as
\begin{align}
  i=&z_0 + 3\;z_1+3^2\;z_2+\dots+3^{\ell-1}\times\underline{0}+\dots\;3^{q-1}\;z_{q-1}\nonumber\\
  j=&z_0 + 3\;z_1+3^2\;z_2+\dots+3^{\ell-1}\times\underline{1}+\dots\;3^{q-1}\;z_{q-1}\nonumber\\
  k=&z_0 + 3\;z_1+3^2\;z_2+\dots+3^{\ell-1}\times\underline{2}+\dots\;3^{q-1}\;z_{q-1}
      \;,
      \label{eq:trit-triples}
\end{align}
where $z_a=0,1,2$, for $a=0,\dots, q-1$. Notice that at layer $\ell$
the members of the triplets, $(i,j,k)$, are numbers that only differ
in the $(\ell-1)$-th trit, while the other $q-1$ trits
$z_a, a\ne \ell-1$, enumerate the $3^{q-1}=n/3$ triplets. (Again, if
more than $q$ layers are needed, we recycle in layer $\ell > q$ the
triplets of layer $\ell\!\!\mod q$.)

Once the triplets of indices, $(i,j,k)$, are selected for each layer,
we can map them onto groups of three indices
$\left(\pi(i),\pi(j),\pi(k)\right)$, via a (randomly chosen)
permutation $\pi$ of the $n$ indices.


The construction above can be generalized for trees of degree $k$, in
which case $k$-tuples of indices can be selected for $k$-qudit gates
to act on.



\subsection*{Two-qudit Inflationary Clifford Gates}
\label{sec:appendix-2-qudit}

In this section we prove:

\vspace{.2cm} 

{\bf Theorem 1}: There exist 2-qudit inflationary
  Clifford gates, for local Hilbert space dimension $d\ge 3$ and $d$
  prime, that expand {\underline{all}} weight-1 generalized Pauli
  strings into weight-2 generalized Pauli strings.

\vspace{.2cm}

We start with a brief review of the higher dimensional Pauli group and
its symplectic representation. Pauli matrices have a natural
generalization in higher dimensions. Define the generalized Pauli
matrices for qudits with local Hilbert-space dimension $d$ (hereafter
assumed to be a prime number) as
\begin{equation}
Z = \sum_{j=0}^{d-1} \omega^j | j \rangle \langle j |, \quad  X = \sum_{j=0}^{d-1} | j \rangle \langle j+1 |,
\end{equation}
where $\omega = e^{i 2\pi /d}$ is the primitive $d$-th root of unity.
The above Pauli operators satisfy the following relations:
\begin{equation}
Z^d = X^d =1, \quad   X Z = \omega ZX.
\end{equation}
It is easy to check that the above matrices reduce to the familiar
Pauli matrices for qubits upon taking $d=2$.

A Pauli string is an element of the Pauli group $\mathcal{P}_n$ acting
on $n$ qudits:
\begin{equation}
Z_1^{u_1}  X_1^{v_1}\otimes  Z_2^{u_2} X_2^{v_2} \otimes \cdots \otimes Z_n^{u_n} X_n^{v_n}, 
\end{equation}
where we have ignored a possible phase factor. The above Pauli string
admits the following symplectic representation as a vector in
$\mathbb{Z}_d^{\otimes 2n}$:
\begin{equation}
g = (u_1, u_2, \ldots, u_n \ |\  v_1, v_2, \ldots, v_n), 
\end{equation}
where $u_i, v_i \in [0, d-1]$.  For $d$ prime, the integers in
$\mathbb{Z}_d$ form a finite field (or Galois field) $\mathbb{F}_d$,
such that the multiplicative inverse exists for each element.  Since
we are interested in the process where a single-site Pauli operator
evolves into a weight-two Pauli string, we focus on $n=2$.  As a
concrete example, Pauli-$Z$ and $-X$ operators acting on site 1 are
represented as vectors:
\begin{equation}
Z_1: \rightarrow g_1 = (1, 0\ | \ 0, 0)  \quad  X_1: \rightarrow g_2 = (0, 0\ | \ 1, 0).
\end{equation}
In the vector representation, products of two Pauli strings correspond
to the addition of the two vectors: $g_1 + g_2$ (mod $d$).

The commutation relation between two Pauli strings in the symplectic
representation can be conveniently computed from the following matrix:
\begin{equation}
\Lambda_{4 \times 4} = 
\begin{pmatrix}
0_{2 \times 2}   &  \mathbb{1}_{2 \times 2}  \\
-\mathbb{1}_{2 \times 2}    & 0_{2 \times 2}
\end{pmatrix},
\end{equation}
namely,
\begin{equation}
\mathcal{S}_1 \mathcal{S}_2 = \omega^r \mathcal{S}_2 \mathcal{S}_1 \   \leftrightarrow  \ g_1 \Lambda g_2^T = r \ ({\rm mod} \ d),
\end{equation}
where $g_1$ and $g_2$ are vectors representing $\mathcal{S}_1$ and
$\mathcal{S}_2$, respectively.

\vspace{.2cm}

\noindent{\bf{Proof of Theorem 1:}} To prove Theorem 1, we need the following lemmas.

\vspace{.2cm}

{\bf {Lemma 1}:} If under a Clifford gate $U_{\rm Cl}$, single-site
Pauli operators $Z_1$ and $X_1$ evolve to weight-2 strings of the form
\begin{equation}
  Z_1: \rightarrow g_1 = (a_1, b_1, 0,0)  \quad X_1:\rightarrow g_2 = (0, 0, \tilde a_1, \tilde b_1)
  \;,
\label{eq:claim1}
\end{equation}
with $a_1, b_1, \tilde a_1, \tilde b_1 = 1, 2, \ldots, d-1$, then all
single-site Pauli operators of the form $Z_1^{u_1} X_1^{v_1}$ will
evolve to weight-2 strings under $U_{\rm Cl}$.

\underline{Proof:} First, we note that both $g_1$ and $g_2$ are
weight-2 strings, as is evident from their vector representations.
Then, consider all other single-site Pauli operators of the form
$Z_1^{u_1} X_1^{v_1}$ with $u_1\neq 0$ and $v_1\neq 0$. Under $U_{\rm
Cl}$, such an operator evolves into
\begin{equation}
Z_1^{u_1} X_1^{v_1}: \rightarrow g = u_1\, g_1 + v_1 \, g_2 \quad ({\rm mod} \ d).
\end{equation}
However, due to properties of the finite field, all elements of $g$
must be nonzero. Hence, we conclude that all single-site Pauli
operators evolve to weight-2 strings under $U_{\rm
Cl}$. $\blacksquare$

In the above lemma, we assume that the Pauli operators $Z$ and $X$ on
site 1 evolve to strings of the form $g_1$ and $g_2$ under a two-qudit
Clifford gate. At this point, it is unclear whether the specific form
of $g_1$ and $g_2$ can be achieved. The answer is affirmative for
$d\geq 3$, as is shown in Lemma 2 below.
\vspace{.2cm}

{\bf Lemma 2}: The form of $g_1$ and $g_2$ in Lemma 1 can always be
achieved via evolution under a Clifford gate $U_{\rm Cl}$ for
$d\geq 3$, while it is not possible for $d=2$.

\underline{Proof:} The only constraint on the time-evolved Pauli strings
$g_1$ and $g_2$ is that they must preserve the commutation relation
$XZ=\omega ZX$ of the original Pauli operators. Using the symplectic
representation, this amounts to the following linear equation:
\begin{equation}
g_1 \ \Lambda \ g_2^T = 1  \quad ({\rm mod} \ d),
\end{equation}
or,  explicitly,
\begin{equation}
a_1\ \tilde a_1 + b_1\ \tilde b_1 = 1 \quad ({\rm mod} \ d).
\label{eq:lemma2}
\end{equation}
For $d\geq 3$, one can take $a_1\tilde a_1>1$ and $b_1\tilde
b_1>1$. Due to properties of the finite field, there always exist
pairs of integers $(x, y)$ in $\mathbb{F}_d$, such that $x+y =1 \
({\rm mod} \ d)$.  We can then take $a_1 \tilde a_1 = x \ ({\rm mod} \
d)$, and $b_1 \tilde b_1 = y \ ({\rm mod} \ d)$.  Again, using
properties of the finite field, it is always possible to find non-zero
$a_1, \ \tilde a_1$, $b_1$ and $\tilde b_1$ that satisfy these two
equations.  Thus, non-zero solutions of Eq.~(\ref{eq:lemma2}) always
exist.

On the other hand, for $d=2$, Eq.~(\ref{eq:lemma2}) can only be
satisfied when $(a_1\tilde a_1, b_1\tilde b_1)=(1,0)$ or
$(0,1)$. Either case implies that one of the four numbers $a_1,
b_1, \tilde a_1, \tilde b_1$ must be zero, which contradicts our
assumption in Lemma 1. In other words, one of the strings $g_1$ and
$g_2$ must have weight 1.  Therefore, the particular form of $g_1$ and
$g_2$ in Lemma 1 cannot be achieved for $d=2$.  $\blacksquare$

Combining the results of Lemma 1 and 2, we have shown that for $d\geq
3$, it is always possible to choose two-qudit Clifford gates such that
all single-site Pauli operators supported on site 1 evolve to weight-2
strings. To complete the proof of Theorem 1, we need to show that the
same Clifford gate is also able to evolve all single-site Pauli
operators on \textit{site} 2 to weight-2 strings.

We show that this is possible by explicitly finding a set of
solutions. We assume that the single-site Pauli operators $Z_1$,
$X_1$, $Z_2$ and $X_2$ evolve into weight-2 Pauli strings of the
following form under $U_{\rm Cl}$:
\begin{eqnarray}
&&Z_1 :\rightarrow g_1 = (1, 1, 0, 0) \quad \quad X_1:\rightarrow g_2=(0,0,\tilde a_1,\tilde b_1)  \nonumber \\
&&Z_2 :\rightarrow g_3 = (a_2, b_2, 0, 0) \quad X_2 :\rightarrow g_4 = (0,0,\tilde a_2, \tilde b_2). \nonumber \\
\end{eqnarray}
Essentially, we have assumed that both $Z_1, X_1$ and $Z_2, X_2$
evolve into the form of Lemma 1, and further take $a_1=b_1=1$. We
demand that the resulting Pauli strings preserve the original
commutation relations, which translates into the following set of
linear equations:
\begin{eqnarray}
\label{eq:1}
\tilde a_1 + \tilde b_1 &=& 1 \quad (g_1\ \Lambda \ g_2^T = 1)\\
\label{eq:2}
a_2 \ \tilde a_2 + b_2 \ \tilde b_2 &=& 1 \quad  (g_3 \ \Lambda \ g_4^T=1)  \\
\label{eq:3}
\tilde a_2 + \tilde b_2 &=& 0  \quad (g_1 \ \Lambda \ g_4^T=0) \\
\label{eq:4}
a_2 \ \tilde a_1 + b_2 \ \tilde b_1 &=& 0 \quad  (g_2\  \Lambda \ g_3^T=0),  
\end{eqnarray}
where the equality mod $d$ is implicit. Notice that with the above
parametrization, the commutation relations $g_1\,\Lambda\, g_3^T=0$
and $g_2\,\Lambda\, g_4^T=0$ are automatically guaranteed.

Our procedure for finding a particular solution to the above set of
equations goes as follows.
\begin{enumerate}
\item We start by solving Eq.~(\ref{eq:1}). Due to properties of the finite field, a solution to Eq.~(\ref{eq:1}) always exists.
\item Next, we solve Eq.~(\ref{eq:3}) by taking $(\tilde a_2, \tilde b_2) = (1, d-1)$.
\item Finally, we find a unique solution $(a_2, b_2)$ by solving
  Eqs.~(\ref{eq:2}) and (\ref{eq:4}).
\end{enumerate}
Of course, the solution is not unique, and we specialize to a
particular one in the above procedure, which suffices to complete the
proof of Theorem 1. $\blacksquare$
\vspace{.2cm}

\noindent{\bf{Examples:}} Below, we give two concrete examples of the construction, for $d=3$
and $d=5$.  \vspace{.3cm}

\underline{$d=3$}: For simplicity, we take Eq.~(\ref{eq:1}) with
$\tilde a_1=\tilde b_1$, which is always possible since the solution
to $2x=1 \ ({\rm mod}\ d)$ always exits for finite fields. We find
$\tilde a_1=\tilde b_1=2$. Next, in step 2, we take $(\tilde
a_2, \tilde b_2) = (1, 2)$. Finally, solving the remaining two
equations yields $(a_2, b_2) = (2,1)$. We thus have
\begin{eqnarray}
&&Z_1 :\rightarrow g_1 = (1, 1, 0, 0) \quad X_1:\rightarrow g_2=(0,0,2,2)  \nonumber \\
&&Z_2 :\rightarrow g_3 = (2, 1, 0, 0) \quad X_2 :\rightarrow g_4 = (0,0,1, 2). \nonumber
\end{eqnarray}

\underline{$d=5$}: Again, we solve Eq.~(\ref{eq:1}) with $\tilde a_1=\tilde b_1$ and
find $\tilde a_1=\tilde b_1=3$. Then, we solve Eq.~(\ref{eq:3}) by
taking $(\tilde a_2, \tilde b_2) = (1, 4)$. Finally, solving the
remaining two equations yields $(a_2, b_2) = (3, 2)$. We thus have
\begin{eqnarray}
&&Z_1 :\rightarrow g_1 = (1, 1, 0, 0) \quad X_1:\rightarrow g_2=(0,0,3,3)  \nonumber \\
&&Z_2 :\rightarrow g_3 = (3, 2, 0, 0) \quad X_2 :\rightarrow g_4 = (0,0,1, 4). \nonumber
\end{eqnarray}


\subsection*{SAC OTOC recursion relations for the three-stage cipher}
\label{sec:appendix-recursion}

\begin{figure}[h!]
\centering
\includegraphics[width=0.3\textwidth]{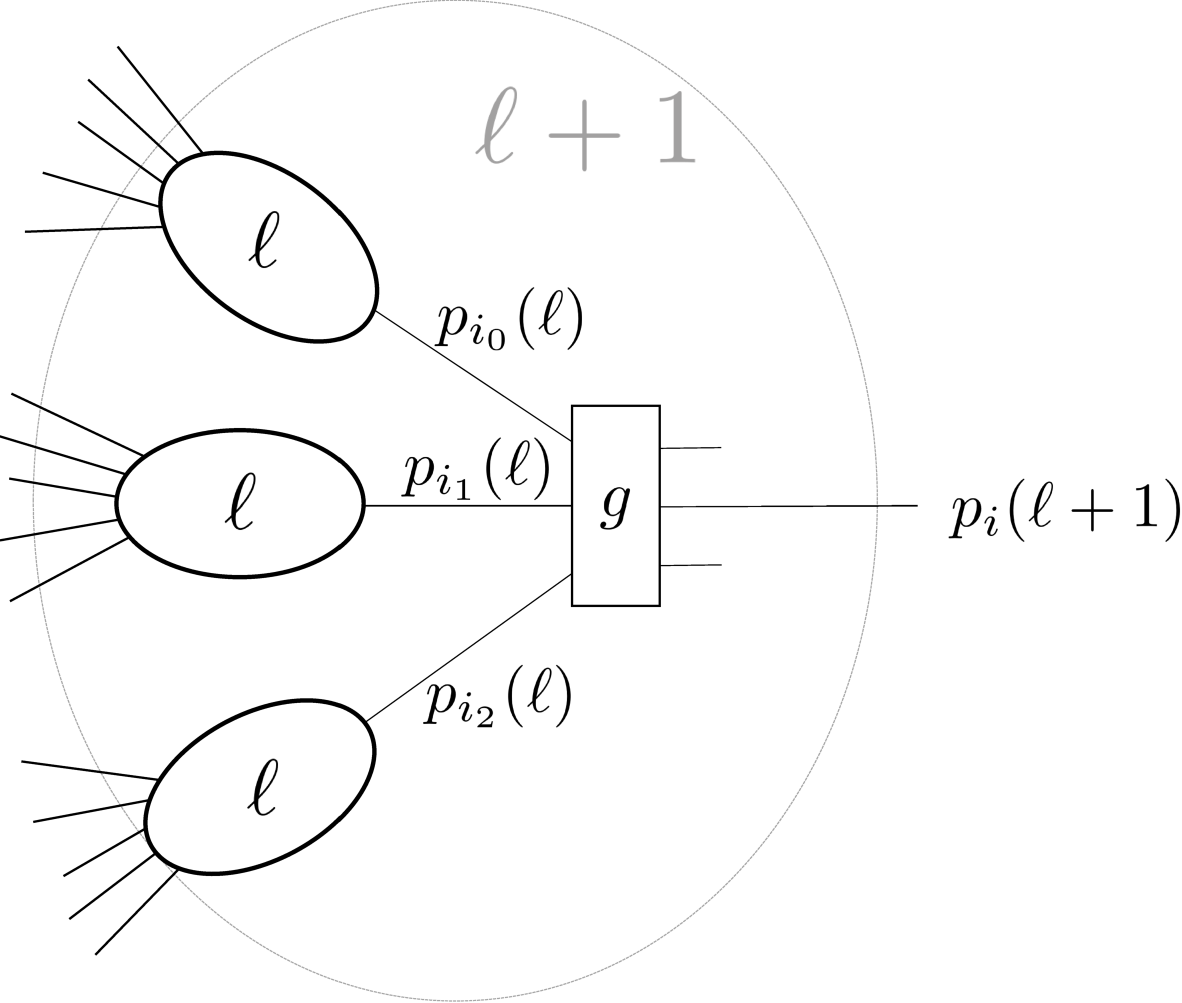}
\caption{{\bf The hierarchical structure of the circuit connectivity.}
  The tree connectivity illustrates the arguments used in the
  derivation of a recursion relation for the probability $p_i$ that a
  bit flips upon flipping a number of inputs.}
\label{fig:tree}
\end{figure}

For the reader's convenience, we reproduce the calculation presented
in Ref.~\cite{cipher-paper} of the square of the SAC OTOC,
$q^{ij}\equiv\left(C^{ij}_{\rm SAC}\right)^2$, as a function of the
number of applied layers of gates $\ell$ of the tree-structured
reversible circuit of 3-bit permutations described in Sec. 7.1 of that
reference.  We use the mean-field assumption (which was checked
numerically in Ref.~\cite{cipher-paper}) that the system
self-averages, implying that $q^{ij}=q$, independent of $i$ and
$j$. The independence of $i$ and $j$ can be traced back to
  the fact that the three bit lines entering the gate $g$ of layer
  $\ell +1$ originate from independent branches of the tree circuit
  emerging from layer $\ell$ (see Fig.~\ref{fig:tree}). As long as
  $\ell \le \log_3 n$, gates in subsequent layers always bring in
  fresh bits and, upon averaging over gates, bitlines $i$ and $j$
  remain uncorrelated.]

We proceed recursively, layer-by-layer, relating $q(\ell+1)$ to
$q(\ell)$.  The calculation is set up in bit space in terms of
probabilities $p_i{(\ell)}$ that after applying the $\ell$-th layer
bit $i$ does not flip.  In the hierarchical tree construction, the
no-flip probability for a given output bit $i$ (at level $\ell+1$) is
determined by the outputs, at bitlines $i_0,i_1, i_2$ coming from
separate branches of the tree (at level $\ell$) and by the 3-bit gate
$g$ in layer $\ell +1$ that takes those three bitlines as inputs and
connects to bit $i$ as one of its outputs.


The specific action of
the gate $g$ determines the fraction of inputs for which the output
$i$ does not flip when $x\to x\oplus c$, with
$x\equiv x_{i_0} + 2\;x_{i_1} + 2^2\,x_{i_2}$ and
$c\equiv c_0+2\;c_1+2^2\,c_2$
encoding which ones of the three bits are flipped ($c_{0,1,2}=0$ for
an unflipped input or 1 for a flipped one). This fraction is expressed
as $C^{g_i}_{c_0c_1c_2}\equiv C^{g_i}_{c}=\left(f^{g_i}_{c}+1\right)/2$, with
\begin{align}
  f^{g_i}_{c}
  =
  \frac{1}{2^3}\sum_{x=0}^7 (-1)^{{g_i}(x)\oplus {g_i}(x\oplus c)}
  \;.
\end{align}
 The recursion for the no-flip probabilities can then be written as
\begin{align}
  \label{eq:app-recursion_pi}
  p_i{(\ell+1)}
  &=
    h\left(p_{i_0}{(\ell)}, p_{i_1}{(\ell)}, p_{i_2}{(\ell)}; \{C^{g_i}_c\}\right)
  \\
  &=
  p_{i_0}{(\ell)}\,p_{i_1}{(\ell)}\,p_{i_2}{(\ell)}\;C^{g_i}_{000}
  +
  (1-p_{i_0}{(\ell)})\,p_{i_1}{(\ell)}\,p_{i_2}{(\ell)}\;C^{g_i}_{100}
  +
    \cdots
    \nonumber\\
  &\;\;\;\;
  +
    (1-p_{i_0}{(\ell)})\,(1-p_{i_1}{(\ell)})\,(1-p_{i_2}{(\ell)})\;C^{g_i}_{111}
    \;.
    \nonumber
\end{align}
We now proceed to consider ensembles of circuits, and analyze the
evolution of the probability distribution, $P(p_i;\ell)$, of the
$p_i$, as function of $\ell$. The recursion relation for
$P(p_i;\ell)$, obtained by using Eq.~\eqref{eq:app-recursion_pi},
reads
\begin{align}
  P{(p_i;\ell+1)}
  &=\sum_{g\in S_8}
    \int dp_{i_0}\,dp_{i_1}\,dp_{i_2}\;
    P(p_{i_0};\ell)\;P(p_{i_1};\ell)\;P(p_{i_2};\ell)\;
    {\cal P}_{\rm set}(g)\;
    \delta\left[p_i-
    h\left(p_{i_0}, p_{i_1}, p_{i_2}; \{C^{g_i}_c\}\right)
    \right]
    \;,
\end{align}
where the gates $g$ are drawn from a probability distribution
${\cal P}_{\rm set}(g)$ that depends on the specific set of gates employed, and which we
assume to be independent of the bitline index $i$. The initial
condition is determined by the fraction $f$ of bits that are
flipped on input:
\begin{align}
  \label{eq:app-initial-P}
  P{(p;\ell=0)}
  =
  f\;\delta(p)+(1-f)\;\delta(p-1)
  \;.
\end{align}
[We note that the
  assumption of independence of the bitline index cannot be justified
  unless $f$ is intensive, which only occurs through the action of
  sufficient number of layers of inflationary gates.]

The evolution of the distribution and the vanishing of the SAC
can be obtained by considering the average and moments of $p$. It is useful to change
variables to $s_i{(\ell+1)}\equiv 2\,p_i{(\ell+1)}-1$, for which the
recursion Eq.~\eqref{eq:app-recursion_pi} reads
\begin{align}
  s_i{(\ell+1)}
  =
  {\widetilde C}^{g_i}_{100}\;s_{i_0}{(\ell)}
  +
  {\widetilde C}^{g_i}_{010}\;s_{i_1}{(\ell)}
  +
  {\widetilde C}^{g_i}_{001}\;s_{i_2}{(\ell)}
  +
  \cdots
  +
  {\widetilde C}^{g_i}_{111}\;s_{i_0}{(\ell)}\;s_{i_1}{(\ell)}\;s_{i_2}{(\ell)}
  \;,
\end{align}
with
\begin{align}
{\widetilde C}^{g_i}_{a}\equiv \frac{1}{2^3}\sum_{c=0}^7 (-1)^{a\cdot c}\;C^{g_i}_{c}
  \;,
\end{align}
where $a\cdot c\equiv a_0\,c_0+a_1\,c_1+a_2\,c_2$.

We are now in position to derive the evolution of the moments
$\overline{s^q{(\ell)}}$. (Even if the distributions for the $s_i$ are
identical, independent of $i$, we keep some of the explicit indices
for bookkeeping of contractions.) The average
\begin{align}
  \overline{s{(\ell+1)}}
  &=
  \sum_{a=1}^7
  \overline{{\widetilde C}^{g_i}_{a}}\;
  \overline{[s_{i_0}{(\ell)}]^{a_0}}\,
  \overline{[s_{i_1}{(\ell)}]^{a_1}}\,
  \overline{[s_{i_2}{(\ell)}]^{a_2}}
    \nonumber\\
  &=
  \sum_{a=1}^7
  \overline{{\widetilde C}^{g_i}_{a}}\;
  \left[\overline{s{(\ell)}}\right]^{a_0+a_1+a_2}\,
  \;.
\end{align}

Similarly, we compute the second moment
\begin{align}
  \overline{s^2{(\ell+1)}}
  =
  \sum_{a,b=1}^7
  \overline{{\widetilde C}^{g_i}_{a}\;{\widetilde C}^{g_i}_{b}}\;
  \overline{[s_{i_0}{(\ell)}]^{a_0+b_0}}\,
  \overline{[s_{i_1}{(\ell)}]^{a_1+b_1}}\,
  \overline{[s_{i_2}{(\ell)}]^{a_2+b_2}}
  \;.
\end{align}

The recursion relations relating $\overline{s}(\ell+1)$ to
$\overline{s}(\ell)$ depend on the gate set used for layer $\ell$
through the coefficients $\overline{{\widetilde C}^{g_i}_{a}}$, which
we present explicitly below for the cases of inflationary and
super-nonlinear gates.  For notational simplicity, we define the
variables $s(\ell)\equiv \overline{s(\ell)}$ and
$q(\ell)\equiv \overline{s^2(\ell)}$.

\begin{figure*}[h]
\centering
\includegraphics[width=0.4\textwidth]{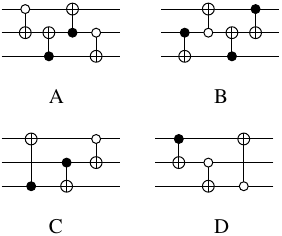}
\caption{{\bf Inflationary 3-bit gates expressed in terms of CNOTs (from
  Ref.~\cite{cipher-paper}).} By permuting bitlines and control
  polarities, one obtains 24 distinct inflationary gates from topology
  A, 24 from B, 48 from C, and 48 from D, for a total of 144.}
  \label{fig:CNOT-equiv-inflationary}
\end{figure*}

\vspace{.2cm}
\noindent{\bf{Inflationary layers:}} Upon computing the averages $\overline{{\widetilde C}^{g_i}_{a}}$ and
$\overline{{\widetilde C}^{g_i}_{a}\;{\widetilde C}^{g_i}_{b}}$ over
the 144 inflationary gates (see
Fig.~\ref{fig:CNOT-equiv-inflationary}), the recursion relations read:
\begin{subequations}
\begin{align}
  \label{eq:app-s-recursion}
  s(\ell+1)
  = \frac{2}{3}\,[s(\ell)]^2 + \frac{1}{3}\,[s(\ell)]^3
  \;,
\end{align}
\begin{align}
  q(\ell+1)
  =
  \frac{2}{3}\,[q(\ell)]^2 + \frac{1}{3}\,[q(\ell)]^3
  \;.
  \label{eq:app-recursion_inflationary_q}
\end{align}
\end{subequations}
The recursion relations of Eqs.~\eqref{eq:app-s-recursion} and
~\eqref{eq:app-recursion_inflationary_q} display two special features:
the first and second moments decouple; and more importantly, the
coefficient of the linear term in $q(\ell)$ in the equation for the
second moment vanishes.

Note that the bimodal initial condition Eq.~\eqref{eq:app-initial-P},
where $p$ only takes values $p=0,1$, implies that an initial $q=1$
cannot evolve under Eq.~\eqref{eq:app-recursion_inflationary_q}, which
displays fixed points at $q=0,1$ (and a non-physical one at $q=-3$).
However, with the deployment of nonlinear gates $q$ drops below 1,
following which inflationary gates significantly accelerate the decay
of $q(\ell)$ with $\ell$ due to the absence of the linear term in
$q(\ell)$ in Eq.~\eqref{eq:app-recursion_inflationary_q}.

\vspace{.2cm}
\noindent{\bf{Super nonlinear layers:}} Using averages $\overline{{\widetilde C}^{g_i}_{a}}$ and
$\overline{{\widetilde C}^{g_i}_{a}\;{\widetilde C}^{g_i}_{b}}$
computed over the 10752 super-nonlinear gates, leads to the recursion
relations characterizing evolution via super nonlinear gates, namely,
\begin{subequations}
\begin{align}
\label{eq:app-supernonlinear_s}
  s(\ell+1)
  =
  \frac{3}{7}\,s(\ell)
  +
  \frac{3}{7}\,[s(\ell)]^2
  +
  \frac{1}{7}\,[s(\ell)]^3
\end{align}
and
\begin{align}
  \label{eq:app-recursion_supernonlinear_q}
  q(\ell+1)
  =&\;
  \frac{3}{28}\,
     \left([s(\ell)]^2 + [s(\ell)]^3\right)
     + \frac{3}{28}\,q(\ell) \,
     \left(1+ 2\,s(\ell) + 2\,[s(\ell)]^2\right)\\
   &+ \frac{3}{28}\,[q(\ell)]^2\,
     \left(1 + s(\ell)\right)
     + \frac{1}{28}\,[q(\ell)]^3
  \;.\nonumber
\end{align}
\end{subequations}
By contrast to the case of inflationary gates, the recursion relations
for $q(\ell+1)$ in Eqs.~\eqref{eq:app-supernonlinear_s} and
~\eqref{eq:app-recursion_supernonlinear_q} depend on both $s(\ell)$
and $q(\ell)$, and contain a term linear in
$q(\ell)$. Eqs.~\eqref{eq:app-s-recursion},~\eqref{eq:app-recursion_inflationary_q},
~\eqref{eq:app-supernonlinear_s}, and
~\eqref{eq:app-recursion_supernonlinear_q} are the starting point for
the discussion of the decay of the SAC OTOC with $\ell$.

\begin{acknowledgments}
  We thank Luowen Qian, Stephen Shenker, and Brian Swingle for useful
  discussions. This work was supported in part by DOE Grant
  DE-FG02-06ER46316 (C.C.), a Grant from the Mass Tech Collaborative
  Innovation Institute (A.E.R.), and a Peking University startup fund and Grant No. 12375027 from the 
  National Natural Science Foundation of China (Z.-C.Y.). C.C. and A.E.R. also
  acknowledge the Quantum Convergence Focused Research Program, funded
  by the Rafik B. Hariri Institute at Boston University.
\end{acknowledgments}


\newpage

\bibliography{references}



\end{document}